\documentclass[aps,twocolumn,prb,showpacs,preprintnumbers,superscriptaddress,amsmath,amssymb]{revtex4-2}

\usepackage[dvips]{graphicx}

\usepackage[english]{babel}
\usepackage{blindtext}
\usepackage{dcolumn}
\usepackage{bm}
\usepackage{ulem}
\usepackage{xcolor}

\begin{document}
\setcounter{secnumdepth}{5}

\title{Terahertz Emission from Mutually Synchronized Stand-Alone Bi$_2$Sr$_2$CaCu$_2$O$_{8+x}$ Intrinsic Josephson Junction Stacks} 

\author{Raphael Wieland}
\thanks{Email: raphael.wieland@uni-tuebingen.de}
\affiliation{Physikalisches Institut, Center for Quantum Science and LISA$^+$, Universit\"{a}t T\"{u}bingen, D-72076 T\"{u}bingen, Germany}

\author{Olcay Kizilaslan}
\affiliation{Physikalisches Institut, Center for Quantum Science and LISA$^+$, Universit\"{a}t T\"{u}bingen, D-72076 T\"{u}bingen, Germany}
\affiliation{Department of Biomedical Engineering, Faculty of Engineering, Inonu University, Malatya, Turkey}

\author{Nickolay Kinev}
\affiliation{Kotel'nikov Institute of Radio Engineering and Electronics, Moscow 125009, Russia}

\author{Eric Dorsch}
\affiliation{Physikalisches Institut, Center for Quantum Science and LISA$^+$, Universit\"{a}t T\"{u}bingen, D-72076 T\"{u}bingen, Germany}

\author{Stefan Guénon}
\affiliation{Physikalisches Institut, Center for Quantum Science and LISA$^+$, Universit\"{a}t T\"{u}bingen, D-72076 T\"{u}bingen, Germany}

\author{Ziyu Song}
\affiliation{Research Institute of Superconductor Electronics, Nanjing University, Nanjing 210023, China}

\author{Zihan Wei}
\affiliation{Purple Mountain Laboratories, Nanjing 211111, China}

\author{Huabing Wang}
\affiliation{Research Institute of Superconductor Electronics, Nanjing University, Nanjing 210023, China}
\affiliation{Purple Mountain Laboratories, Nanjing 211111, China}

\author{Peiheng Wu}
\affiliation{Research Institute of Superconductor Electronics, Nanjing University, Nanjing 210023, China}
\affiliation{Purple Mountain Laboratories, Nanjing 211111, China}

\author{Dieter Koelle}
\affiliation{Physikalisches Institut, Center for Quantum Science and LISA$^+$, Universit\"{a}t T\"{u}bingen, D-72076 T\"{u}bingen, Germany}

\author{Valery P. Koshelets}
\affiliation{Kotel'nikov Institute of Radio Engineering and Electronics, Moscow 125009, Russia}

\author{Reinhold Kleiner}
\thanks{Email: kleiner@uni-tuebingen.de}
\affiliation{Physikalisches Institut, Center for Quantum Science and LISA$^+$, Universit\"{a}t T\"{u}bingen, D-72076 T\"{u}bingen, Germany}

\date{\today}

\begin{abstract}
Suitably patterned single crystals made of the cuprate superconductor Bi$_2$Sr$_2$CaCu$_2$O$_{8+x}$ (BSCCO), intrinsically forming  a stack of Josephson junctions, can generate electromagnetic radiation in the lower terahertz regime. Due to Joule heating the emission power of single stacks seems to be limited to values below 100\,$\mu$W. To increase the radiation power, mutually synchronized arrays situated on the same BSCCO base crystal have been studied. A maximum power of almost 1\,mW has been achieved by synchronizing three stacks. Mutual electromagnetic interactions via a connecting BSCCO base crystal have been considered essential for synchronization, but the approach still suffers from Joule heating, preventing the synchronization of more than three stacks. In the present paper we show, on the basis of two emitting stacks, that mutual synchronization can also be achieved by stand-alone stacks contacted by gold layers and sharing only a common gold layer. Compared to BSCCO base crystals, the gold layers have a much higher thermal conductivity and their patterning is not very problematic. We analyze our results in detail, showing that the two oscillators exhibit phase correlations over a range of $\pm$0.4\,GHz relative to their center frequencies, which we mainly studied between 745\,GHz and 765\,GHz. However, we also find that strong phase gradients in the beams radiated from both the mutually locked stacks and the unlocked ones play an important role and, presumably, diminish the detected emission power due to destructive interference. We speculate that the effect arises from higher-order cavity modes which are excited in the individual stacks. Our main message is that the mutual interaction provided by a common gold layer may open new possibilities for relaxing the Joule-heating-problem, allowing the synchronization of a higher number of stacks. The approach may also allow one to synchronize several stacks, which are comparatively small in size and less prone to the strong phase gradients we observed. Our findings may boost attempts to substantially increase the output power levels of the BSCCO terahertz oscillators.

\end{abstract}

\maketitle

\section{Introduction}
\label{sec:intro}

The cuprate superconductor Bi$_{2}$Sr$_{2}$CaCu$_{2}$O$_{8+x}$ (BSCCO), once properly patterned, is known to emit electromagnetic radiation in the range from 0.2 up to a few terahertz (THz). This frequency regime is highly interesting for applications but only sparsely populated with compact solid state sources \cite{Ferguson02, Tonouchi07,Williams07,Feiginov14}. 
BSCCO is a layered superconductor with alternating superconducting and insulating sheets. A single crystal thus forms a natural stack of intrinsic Josephson junctions (IJJs), with $\sim$670 junctions per $\mu$m of crystal thickness \cite{Kleiner92}.
In the resistive state the supercurrents across each IJJ oscillate with a frequency $f_{\rm J} = V_{\rm J}/\Phi_0$, where $V_{\rm J}$ is the voltage across the junction and $\Phi_0$ is the flux quantum, $\Phi_0^{-1}$ = 483.6\,GHz/mV. Provided, that the IJJs can be phase-synchronized, such a stack can act as a voltage-tunable emitter of coherent electromagnetic waves. BSCCO emitters have attracted great interest in recent years, 
in terms of both experiment \cite{Ozyuzer07,Wang09a,Minami09,Kurter10,Wang10a,Tsujimoto10,Orita10,Benseman11,Yamaki11,
Li12,Kakeya12,An13,Benseman13,Benseman13a,Minami14,Ji14,Kashiwagi15b,Zhou15a,Zhou15b,Watanabe15,Kashiwagi15a,Kashiwagi15c,Gross15,Benseman15,Nakade16,Sun17,Elarabi17, Borodianskyi17, Kashiwagi18,Minami19,Zhang19,Benseman19,Kuwano20,Tsujimoto20,Ono20,
Catteano21,Tsujimoto21,Kuwano21,Kashiwagi21,Sun23,Kihlstrom23,Miyamoto24,Elarabi24} 
and theory \cite{Bulaevskii07,Lin08,Hu08,Klemm09,Krasnov10,Koshelev10,Kadowaki10,Yurgens11,Lin12,Asai12,Asai12b,Gross12,Lin13,Rudau15, Rudau16, Asai17,Cerkoney17, Klemm17b, Rain21,Krasnov21,Kalhor21,Kobayashi22,Kobayashi22b}.
For recent reviews, see Refs. \onlinecite{Welp13,Kakeya16,Kashiwagi17,Kleiner19,Delfanazari20}.
Coherent off-chip terahertz emission in the frequency range between 0.5 and 0.85\,THz was first reported for rectangular and 1-$\mu$m-thick mesas patterned on a BSCCO base crystal, with an extrapolated output power of up to 0.5\,$\mu$W \cite{Ozyuzer07}. The mesas were about 300\,$\mu$m long and several tens of $\mu$m wide.
The emission frequency was found to be inversely proportional to the width of the stack, leading to the conclusion that resonant cavity modes oscillating along the width of the stack play an important role in synchronizing the junctions in the stack. Later on, a variety of cavity resonances have indeed been found and analyzed \cite{Lin08,Minami09,Wang09a,Kadowaki10,Wang10a,Koshelev10,Tsujimoto10,Cerkoney17,Kashiwagi18,Zhang19,Benseman19,Tsujimoto16,Sun23}.

In addition to the mesa-type structures, IJJ stacks have also been realized as stand-alone structures, where the BSCCO stack is contacted by gold layers from both sides, and sometimes also as z-type structures where the stack plus contacting electrodes were patterned from a solid BSCCO single crystal \cite{Kleiner19,Gross15}.
It further turned out for all structures that Joule heating plays an important role and for a large input power leads to the formation of a hot spot, a region within the stack with a temperature above the critical temperature $T_{\rm c}$  \cite{ Wang09a,Wang10a,Yurgens11,Gross12,Kakeya12,Benseman13,Minami14}.
The hot spot can coexist with regions that are still superconducting and produce terahertz radiation. 
Joule heating is also reflected in the shape of the current-voltage characteristic (IVC) of the stack. For low bias currents the  input power is moderate and the temperature distribution in the stack is almost homogeneous and close to the bath temperature $T_{\rm{bath}}$. In this ``low-bias regime'' the IVCs exhibit a positive differential resistance. With increasing current, due to the fact that the out-of-plane resistivity of BSCCO decreases with increasing temperature, the IVCs start to back-bend and above some threshold current the hot spot forms and increases in size with increasing current and input power (``high-bias regime'').  Joule heating and the presence of a hot spot affects radiation. On the one hand it is found that the linewidth of radiation is much lower in the presence of the hot spot \cite{Li12}, and in addition the emission frequency becomes tunable by manipulating the hot spot's size and position \cite{Zhou15a,Zhou15b, Watanabe15}. On the other hand, Joule heating limits the maximum voltage across the junctions and thus the maximum emission frequency. Joule heating also limits the total number of junctions $N$ in the stack and thus the maximum emission output power, which ideally should scale $\propto N^2$.
Thermal management thus became an issue over the years. Originally, the IJJ stacks were realized as mesa structures patterned on BSCCO single crystals that were just glued with epoxy to a cooled substrate \cite{Ozyuzer07}. To improve cooling, the crystals have e.g. been soldered to Cu substrates \cite{Benseman15}, and double-sided cooling techniques have been applied to stand-alone stacks \cite{Ji14, Kashiwagi15a, Kashiwagi15b, Kashiwagi15c}. This way, output powers of several tens of $\mu$W have been achieved. The maximum emission frequencies increased to 2.4\,THz and $N$ increased to more than 3000.

Even after improvements the maximum emission frequencies seem to be limited to values below 2.5\,THz and the maximum emission power to values below 100\,$\mu$W for the large structures described above. 
In terms of emission frequencies, substantially higher values of up to 11\,THz have been obtained by using stacks with much smaller in-plane dimensions of around 10\,$\mu$m consisting of only around 200 IJJs \cite{Borodianskyi17}. The radiation power efficiency for the typically used IJJ stacks is presently well below 1$\%$, perhaps with one exception \cite{Catteano21}. To increase this efficiency, the use of properly designed antenna structures \cite{Asai12b,Minami19,Kashiwagi21,Kuwano21,Tsujimoto21,Rain21,Kalhor21,Krasnov21,Krasnov23}, or the use of resonators \cite{Ono20} has been proposed and partially realized. 

The third approach, which can presumably be combined with the former ones, is to mutually phase-lock planar arrays of stacks located on the same chip, like for arrays of conventional Josephson junctions \cite{Benz91,Barbara99,Darula99,Song09,Galin20,Galin22}. The present paper intends to contribute to this approach.
Criteria to distinguish locked states from unlocked ones can be the narrowing of the linewidth of radiation, the increase of the emitted radiation power, the comparison of the dc voltages across the stacks, or the change in the angular dependence of the polarization of the emitted radiation patterns. 
Presently, the amount of literature on coupled IJJ stacks is modest \cite{Orita10,Benseman13a,Lin13,Gross15,Tsujimoto20,Kobayashi22,Kobayashi22b,Koshelets19,Lin14}. In Ref. \onlinecite{Gross15} the mutual interaction of two z-type structures was investigated, but no clear signatures of mutual phase-lock were found. Refs. \onlinecite{Orita10,Benseman13a,Lin13,Tsujimoto20,Kobayashi22,Kobayashi22b} considered mesa-type structures. Clear evidence for mutual phase-lock was given and it was concluded that the common base crystal is required to provide an electromagnetic interaction via Josephson plasma waves propagating in the base crystal. In particular, Ref. \onlinecite{Benseman13a} reported an output power of up to 0.63\,mW for an emission frequency of 0.51\,THz when synchronizing 3 adjacent mesas, with spacing of 60\,$\mu$m between the mesas. The main obstacle to mutually synchronize more stacks is still Joule heating, which is reinforced by the poor thermal conductance of the BSCCO base crystal but, on the other hand, seems to be necessary for mutual coupling. Below, we report on the synchronization of two stand-alone stacks which share a common Au base electrode, with a spacing between the stacks of 200\,$\mu$m.  The data show that a common base crystal and thus an interaction via Josephson plasma waves is not required for mutual electric coupling. This observation may indicate a possibility to relax the notorious heating problem to synchronize a larger number of stacks. 
\\

\section{Samples and experimental techniques}
\label{sec:samples}

The sample investigated is patterned from a slightly underdoped BSCCO single crystal, with a $T_{\rm c}$ of 86\,K. The stacks are realized as gold-BSCCO-gold (GBG) structures, using a procedure similar as the one described in \cite{Ji14}. In brief, a thin flake of the crystal is glued by epoxy resin onto a 1-mm-thick sapphire substrate. The upper layers are removed using scotch tape and the new crystal surface is covered by about 200\,nm of gold (thermal evaporation). Using optical lithography and Argon-ion etching five rectangular mesas are patterned onto the gold covered BSCCO. Another sapphire substrate is glued onto them, then the first substrate is removed, cleaving off the common base crystal. Another gold layer with a thickness of 70\,nm is evaporated onto the fresh BSCCO surface.

\begin{figure}[h]
	\includegraphics[width=\columnwidth,clip]{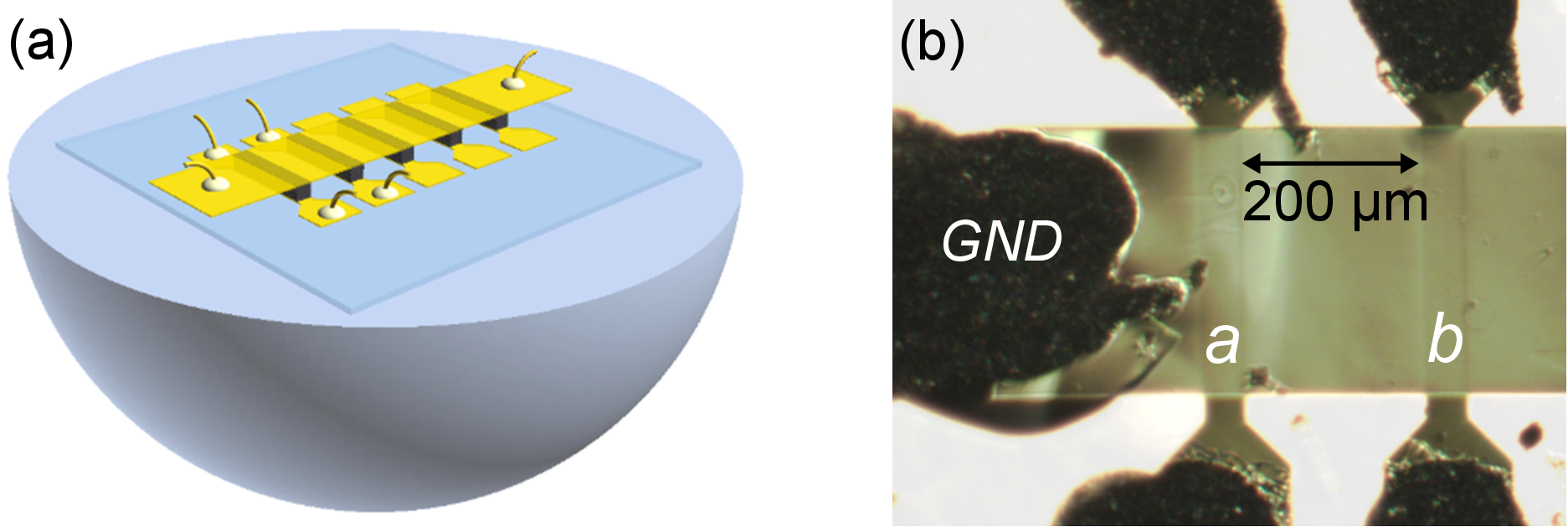}
	\caption{Sample geometry. (a) Schematic of sample mounted to hemispheric lens. (b) Optical image of stacks\;$a$ and $b$ used for synchronization experiments. 
	}
	\label{fig:sample}
\end{figure}

By optical lithography, followed by ion etching, the width of this film was reduced to 300\,$\mu$m along the long side of the IJJ stacks, the latter protruding about 150\,$\mu$m on both sides from the gold. These parts are etched down until the lower gold layer is accessible. The remaining parts of the lower gold layer are used to individually make electrical contact to the stacks, while the upper, 70-nm-thick gold layer, still stretching across all stacks, is used as common ground. A sketch of the completely fabricated sample is shown in Figure \ref{fig:sample}(a). Two of the stacks, denoted stack\;$a$ and stack\;$b$, were used for the synchronization experiments. An optical image of these stacks is shown in Figure \ref{fig:sample}(b). The stacks’ dimensions are approximately 50 $\times$ 300 $\times$ 0.7\,$\mu \rm m^3$, and the number of junctions per stack is $N \approx 460$, as determined from emission measurements. The gap between the stacks is 200\,$\mu$m in width.

For the described sample design each stack has two BSCCO-gold interfaces, one on the top side and one on the bottom side, allowing for a 2-terminal scheme for measurements of the IVC of each stack, which includes the contact resistances between BSCCO and the gold layers.  For emission measurements the sample is mounted on a hemispheric sapphire lens, as sketched in Figure \ref{fig:sample}(a).

\begin{figure}[h]
	\includegraphics[width=\columnwidth,clip]{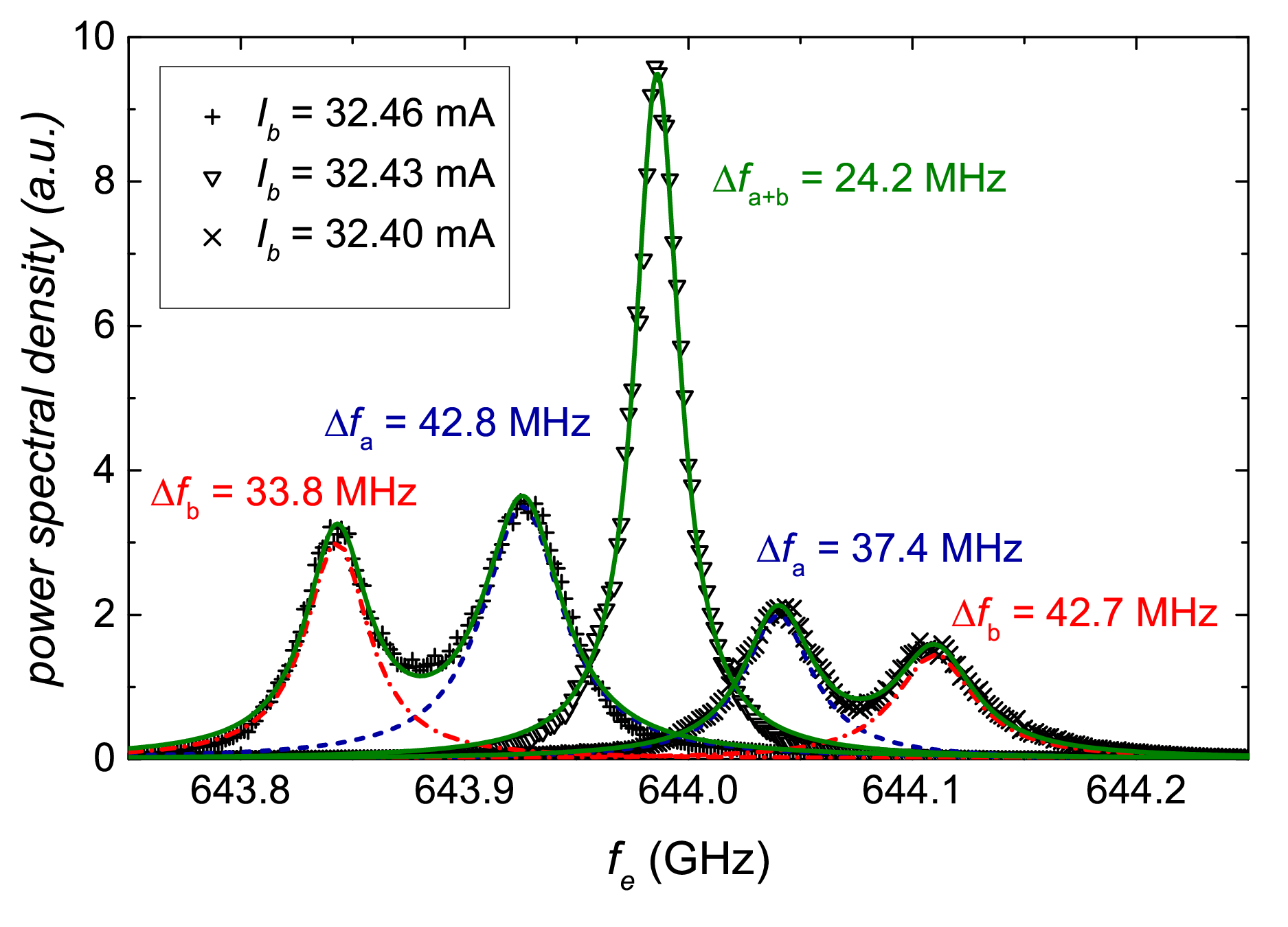}
	\caption{
		Selected emission data obtained with the SIR: Power spectral density for three different current values of stack\;$b$, 32.46\,mA (+), 32.43\,mA ($\nabla$) and  32.40\,mA ($\times$). The current through stack\;$a$ remained constant, $I_a = $ 24.24\,mA. Data are fitted by Lorentzians. Two Lorentzians have been used for, respectively, $I_b = $ 32.46\,mA and 32.40\,mA. Their sum is plotted in solid green, the one attributed to stack\;$a$ in dashed blue, the one for stack\;$b$ in dash-dotted red. A single Lorentzian has been used for $I_b = $ 32.43\,mA, plotted in solid green. The extracted linewidths of radiation are indicated.
	}
	\label{fig:SIR}
\end{figure}

\begin{figure}[h]
	\includegraphics[width=\columnwidth,clip]{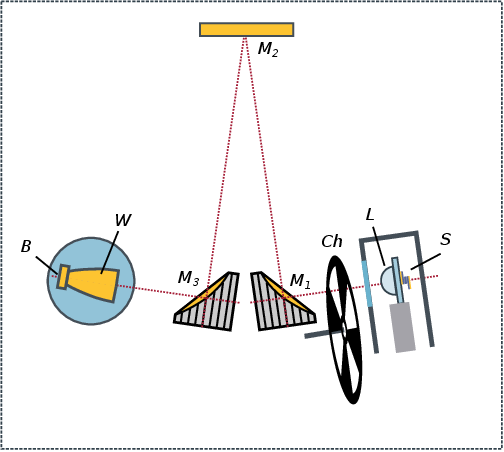}
	\caption{Geometry of the experiment. The sample $S$ is mounted on a hemispheric lens $L$, The emitted light fields propagate (roughly) along the optical path indicated by the dotted line. Having left the cryostat the light fields get modulated by the chopper $Ch$, are reflected by the first parabolic mirror $M_1$ and then pass the planar mirror $M_2$ and the second parabolic mirror $M_3$. They are subsequently collected by the Winston cone $W$ and directed to the bolometer $B$, which produces a time-modulated intensity which is subsequently processed by a vector lock-in amplifier.    
	}
	\label{fig:setupTue}
\end{figure}

The sample was pre-characterized by transport and emission experiments using the niobium-based superconducting receiver (SIR) as a detector for the emitted radiation, giving evidence that the two stacks can be phase-locked. Some results and details of the setup are reported in \cite{Koshelets19}. In brief, the setup makes use of two optical Helium bath cryostats, one hosting the emitter and the other one the detector. The incoming signal at a frequency $f_{\mathrm{s}}$ is mixed with the reference signal of a superconducting Josephson junction local oscillator at frequency $f_{\mathrm{LO}}$ to yield a difference (intermediate) frequency $f_{\mathrm{IF}}$ which is analyzed conventionally in subsequent steps \cite{Koshelets15}. The frequency resolution of this heterodyne detection scheme is better than 1\,MHz and required to resolve the linewidth of radiation of the BSCCO emitter. With the local oscillator fixed at 650\,GHz and the unregulated sample holder kept at 14\,K, phase synchronization of the two stacks was observed when varying the bias current of one of the stacks while keeping the bias current through the other stack fixed. Figure \ref{fig:SIR} shows some additional results. In the unlocked regime two emission lines with individual linewidths of around 35--40\,MHz were visible. In the locked regime the two lines collapsed to a single line with an emission frequency near 644\,GHz and a linewidth of about 24\,MHz. Also, the integrated emission power exceeded the sum of the individual emission powers by 15--20\,\%.
After 2021 we continued our study of mutual synchronization using the setup shown schematically in Figure \ref{fig:setupTue}. Here we have much less frequency resolution as in the SIR-setup but it is possible to cover a wider range of emission frequencies. Also, with more data points measured, it is easier to distinguish between power modulations of the individual stacks and and power changes due to mutual synchronization. 

In our setup the sample $S$ consisting of the two stacks is mounted near but not exactly in the focus of the hemispheric lens $L$. The wiring from the sample through the cryostat is done using separate wires for current and voltage (for the top contact of stack\;$b$: only one wire). After having passed the lens, the beams emitted by the stacks have presumably widened to the mm scale. As shown previously \cite{Tsujimoto16} the intensity of radiation emitted by one stack after the lens is strongly peaked in forward direction, with a divergence of around $\pm$5\textdegree. Since the position of the two stacks is somewhat out of the focal point of the lens, the two radiation beams are likely to be directed slightly away from the optical axis. Next, the emission fields are directed through the HDPE-window of the optical He-flow cryostat and pass the plane of the chopper $Ch$, which periodically opens and closes with a frequency $\omega_{Ch }=2\pi\cdot$ 80\,Hz. The chopper is located at a distance $\sim$ 6\,cm from the lens. In the plane of the chopper the beams emitted by the two stacks have widened to a diameter of around 2\,cm and will partially overlap. After the chopper the beams first get reflected by a 90\textdegree\, parabolic mirror $M_1$, then by the planar mirror $M_2$, followed by a second 90\textdegree\, parabolic mirror $M_3$ which focuses the beams into the entrance of the Winston cone $W$ located in a He bath cryostat. The Winston cone guides the signal, through several reflections, to the Ge-bolometer $B$ mounted on the exit side of the Winston cone. The bolometer detects the signal incoherently. Finally, the signal created by the bolometer, which is periodically modulated in time by the chopper, is multiplied by the reference phase of a vector lock-in amplifier and integrated over time. The lock-in amplifier is set to record the bolometer signal dominantly in channel X and for a smaller part in channel Y.  

The flat mirror $M_2$ can be replaced by a home-made lamellar split mirror for Fourier spectroscopy, with a frequency resolution of about 10\,GHz \cite{Guenon10}.  This interferometer can be used for a first but not very precise determination of the emission frequency.

Usually, the whole interferometer setup is flooded with nitrogen gas to reduce water vapor absorption. However, in some experiments discussed below we flood with air and use the water vapor absorption line at  $f_{\mathrm{water}} \approx$\,752.033\,GHz for frequency calibration. This is necessary for a precise comparison of the emission frequency of the two stacks.

\section{Results}
\label{sec:results}

\begin{figure}[h]
	\includegraphics[width=\columnwidth,clip]{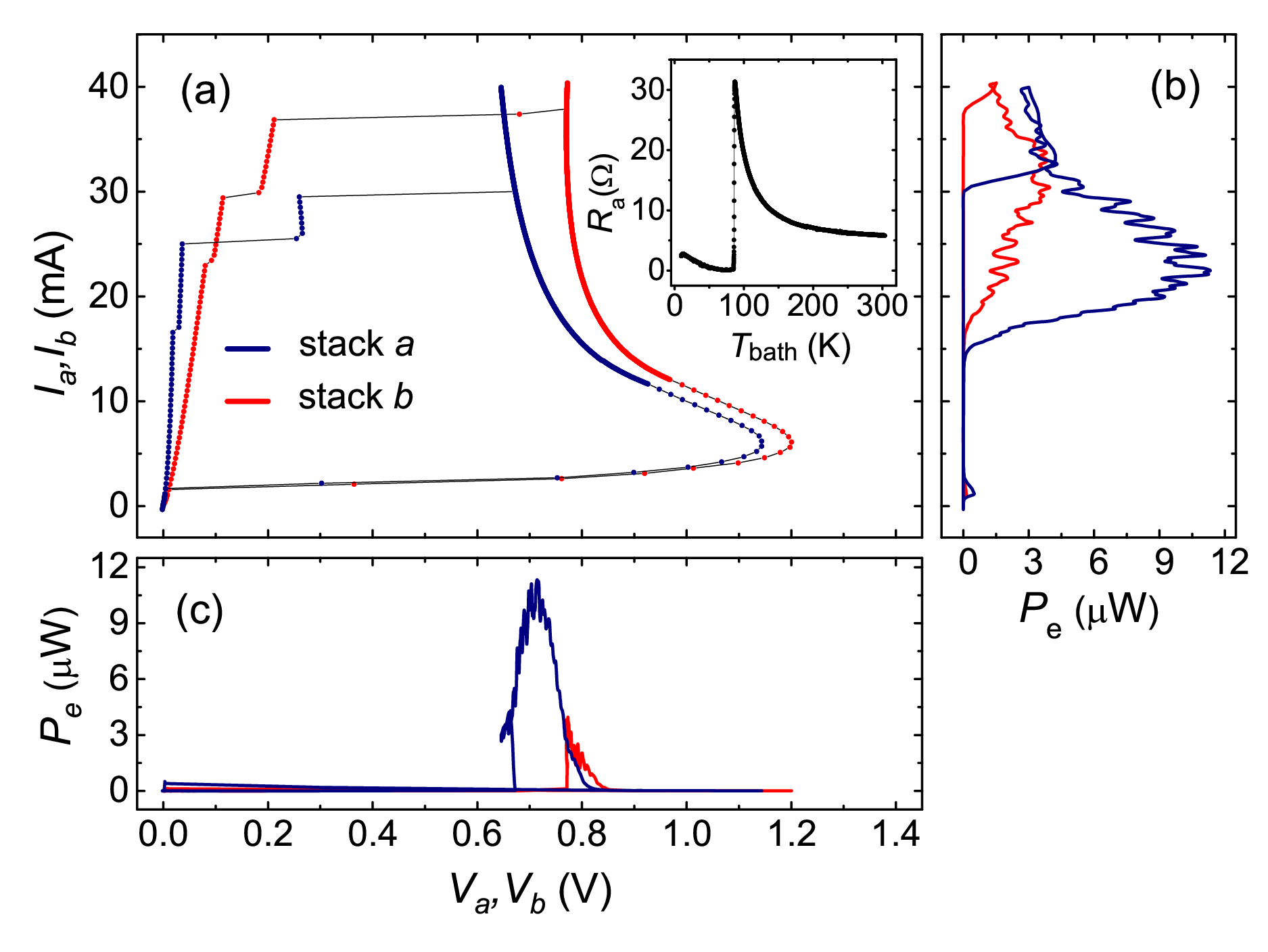}
	\caption{(a) IVCs of stack\;$a$ and $b$, biased only one at a time, at $T_{\mathrm{bath}}=$\,10\,K. Inset shows resistance vs. bath temperature for stack\;$a$. (b),(c) emission power, as detected by the bolometer as function of, respectively, bias current (b) and voltage (c) across the stacks. Emission measurements were done simultaneously with the corresponding IVC measurement. 
	}
	\label{fig:IVC}
\end{figure}

As a pre-characterization the inset of Figure \ref{fig:IVC}(a) shows the resistance of stack\;$a$ vs. bath temperature  $T_{\rm{bath}}$. The resistive transition occurs at $T_{\rm c} \approx$\,86\,K and the overall shape of $R_{\rm a}$ vs. $T_{\rm{bath}}$ indicates that the sample is slightly underdoped \cite{Watanabe97}. The main panel of Figure \ref{fig:IVC}(a) shows the IVCs of the two stacks at $T_{\rm{bath}}=10$\,K. The stacks were biased only one at a time. For each stack, the current is increased from 0 to 40\,mA and subsequently reduced back to zero. The voltage appears as measured, i.e., contact resistances are included. One observes the typical hysteretic IVCs of IJJ stacks, where upon increasing current, groups of IJJs successively get resistive. For $I_{\rm a} > $\,30\,mA ($I_{\rm b} > $\,37 mA) all IJJs within stack\;$a$ (stack\;$b$) are resistive and remain in this state until the current is reduced to values below 2\,mA. Figures \ref{fig:IVC}(b) and (c) show the emitted power as a function of, respectively, current (b) and voltage (c), as detected by the bolometer while recording the IVCs. For both stacks, emission mainly occurred in the high-bias regime, where a hot-spot has formed inside the stacks. Maximum emission power readings were around 10\,$\mu$W for stack\;$a$ and 4\,$\mu$W for stack\;$b$. 

Note that the bolometer readings for both stacks show oscillations both as a function of current and as a function of voltage. This actually translates into a modulated emission power as a function of emission frequency, where the quasi-periodic oscillations occur on a frequency scale of around 3\,GHz up to some 10\,GHz. In fact, many published data show this effect (\cite{Tsujimoto14,Zhou15b,Rudau15,Rudau16,Kobayashi22b}). The modulations could, in principle, arise from the excitation of different cavity modes in the stack. The effect is well known and has been analyzed intensively in the literature, see e.g. \cite{Tsujimoto16,Sun23}. However, the resonance frequencies of different cavity modes are separated by tens of GHz, and there are simply not enough cavity resonances to explain the numerous oscillations we observe. In addition, we tried to correlate the linewidth of radiation, as measured by the SIR, with the few-GHz-modulations. The result is negative, we could not see any correlation.  Thus, we conclude that these oscillations are due to interference which occurs outside of the stacks. Let us assume that after passing the lens the different parts of the coherent beam (of a single stack) with an initially smooth phase profile have travelled a path difference $\Delta l$ before the detection by the bolometer. Then, one would expect a phase difference of 2$\pi$ between the two paths for a frequency difference $\Delta f$ of order $c/\Delta l$, yielding  $\Delta l =$\,10\,cm for $\Delta f =$\,3\,GHz. This value is by far too large to be explainable by inaccuracies in our detection scheme, e.g. by geometric aberrations caused by the parabolic mirrors \cite{Chopra23} or by multiple reflections in the Winston cone. We have also ruled out that the cryostat windows or reflections inside the cryostat are the origin of the modulations. A remaining possibility arises from the fact that for a given cavity mode the Josephson phase along the edges of the stack can vary strongly. Below we give arguments that a (1,3) cavity mode has been excited in our experiments. Here, the resonance pattern forms one half-wave along the width and 3 half-waves along the length of the stack, thus the phase gradients amounts to 3$\pi$ along the length and $\pi$ along the width. The lens may heavily mix the corresponding wave fronts. This may also change the directivity of the emitted radiation when varying the emission frequency. Whether or not this effect can explain the observed oscillations is unclear. We thus need to treat the origin of the power oscillations and the phase profiles of the emitted radiation beams as an unknown when analyzing the interference of the light fields created by the two oscillators.

In the experiments discussed next we perform a broadband measurement of the combined emission power of both stacks while varying their bias currents $I_{\rm a}$ and $I_{\rm b}$ over some range.
 
\begin{figure}[h]
	\includegraphics[width=\columnwidth,clip]{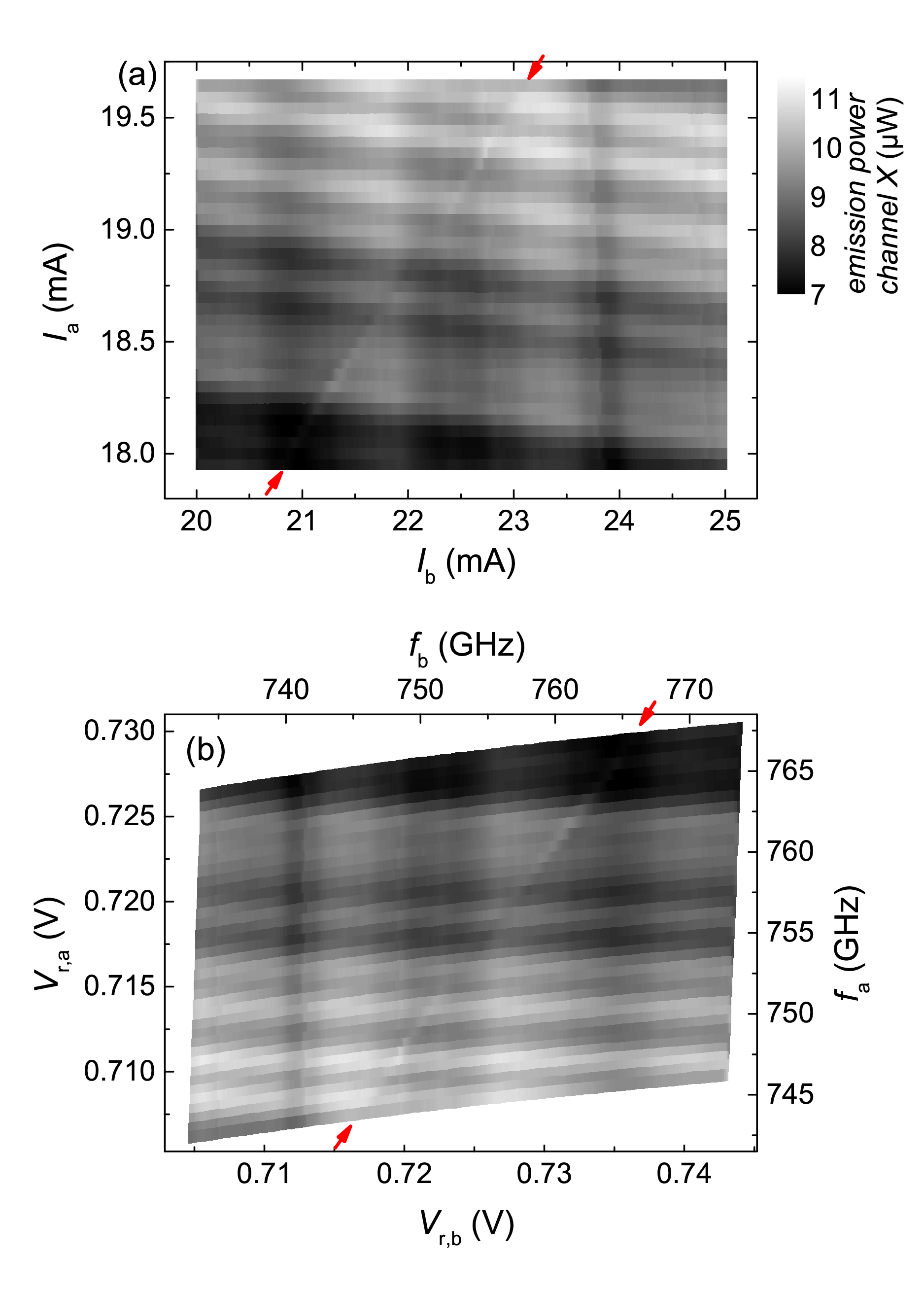}
	\caption{For lock-in channel X: Emission data for the two stacks biased at the same time. (a) Bolometric power as grayscale plotted against stack currents $I_{\rm a}$ and $I_{\rm b}$. (b) Bolometric data as a function of the rescaled voltages $V_{\rm{r,a}}$ and $V_{\rm{r,b}}$ (left and bottom axes) and the corresponding emission frequencies $f_{\rm a}$ and $f_{\rm b}$ (right and top axes). For determining $V_{\rm{r,a}}$ and $V_{\rm{r,b}}$ the voltages arising from contact resistances $R_{\rm a}=$\,0.5\,$\Omega-$\,4\,$\Delta I_{\rm b}$\,$\Omega$/A and $R_{\rm b}=$\,2.8\,$\Omega$ were subtracted from the measured voltages. Here, $\Delta I_{\rm b}=I_{\rm b}-$\,20\,mA.  To convert to emission frequencies junction numbers $N_{\rm a}=$\,460 for stack\;$a$ and $N_{\rm b}=$\,465 for stack\;$b$ were used. Every data point is represented by a rectangle of reduced aspect ratio to accommodate for the higher density of points along $I_{\rm b}$ and $V_{\rm{r,b}}$.   
	}
	\label{fig:tartanRawX}
\end{figure}

The result is shown in Figure \ref{fig:tartanRawX}(a) for lock-in channel X. One notes oscillations of the detected power slightly tilted from the vertical and horizontal directions, plus a strongly tilted ``diagonal'' line. The slightly tilted lines resembling a Scottish tartan arise from the power modulations of the individual stacks, as already visible in Figure \ref{fig:IVC}(b). In general we may write the total detected power as
\begin{equation}
	\label{eq:totalPowerCurrent}
	\begin{split}
	P_{\rm{tot,\alpha}}(I_{\rm a},I_{\rm b}) &= P_{\rm{a,\alpha}}(I_{\rm a}) + P_{\rm{b,\alpha}}(I_{\rm b}) \\
	&+ P_{\rm{ab,\alpha}}(I_{\rm a},I_{\rm b}) .
	\end{split}
\end{equation}
$P_{\rm{a,\alpha}}(I_{\rm a})$ and $P_{\rm{b,\alpha}}(I_{\rm b})$ are the emission powers of the individual stacks if they were independent of each other. The index $\alpha$ indicates whether the signal is detected in channel X or Y, i.e., $\alpha$ equals either X or Y.  $P_{\rm{ab,\alpha}}(I_{\rm a},I_{\rm b})$ stands for the additional emission when the stacks are interacting.  $P_{\rm{a,\alpha}}(I_{\rm a}) + P_{\rm{b,\alpha}}(I_{\rm b})$ constitute the tartan pattern. The feature of interest is the diagonal line indicating interaction represented by $P_{\rm{ab,\alpha}}(I_{\rm a},I_{\rm b})$, which basically represents the interference between the two oscillators.

In order to convert the current scales to the more interesting voltage and frequency scales we need to have a good estimate of the number of IJJs in the stack, and we need to properly subtract the voltages arising from contact resistances from the measured voltages $V_{\rm a}$ and $V_{\rm b}$. To find the contact resistances we make use of the fact that the tartan-like oscillations should appear vertical and horizontal when the combined signal of stacks $a$ and $b$ is plotted as a function of the bare voltages $V_{\rm{r,a}}$ and $V_{\rm{r,b}}$ after subtration of the contact resistances. Details are given in Appendix \ref{Appendix A}. For the contact resistances near the voltages of interest we find $R_{\rm a}=0.5\,\Omega-4 \Delta I_{\rm b}\Omega/\rm{A}$ for stack\;$a$ and $R_{\rm b}=2.8\,\Omega$ for stack\;$b$. The term 4\,$\Delta I_{\rm b}$\,$\Omega$/A, with $\Delta I_{\rm b}=I_{\rm b}-20$\,mA is induced by the Joule power dissipated in stack\;$b$. Figure \ref{fig:tartanRawX}(b) shows the resulting data for the bare voltages $V_{\rm{r,a}} = V_{\rm a}-R_{\rm a}I_{\rm a}$ and $V_{\rm{r,b}} = V_{\rm b}-R_{\rm b}I_{\rm b}$ . To find the junction numbers in stacks\;$a$ and $b$ we make use of the 752.033\,GHz water absorption line which becomes visible when the interferometer setup is flooded with air. The procedure is outlined in detail in Appendix \ref{Appendix A}. For the junction numbers we obtain $N_{\rm a}=$\,460$\pm$1 for stack\;$a$ and $N_{\rm b}=$\,465$\pm$1 for stack\;$b$. 

For lock-in channel X the result of the conversion is shown in Figure \ref{fig:tartanRawX}(b). In the ($V_{\rm{r,a}}$,  $V_{\rm{r,b}}$) or, respectively the  ($f_{\rm a}$,  $f_{\rm b}$) plane the tartan-like pattern forms vertical and horizontal stripes and the ``diagonal'' line appears when $f_{\rm a}$ and  $f_{\rm b}$ (nearly) coincide.

\begin{figure*}
	\includegraphics[width=\textwidth,clip]{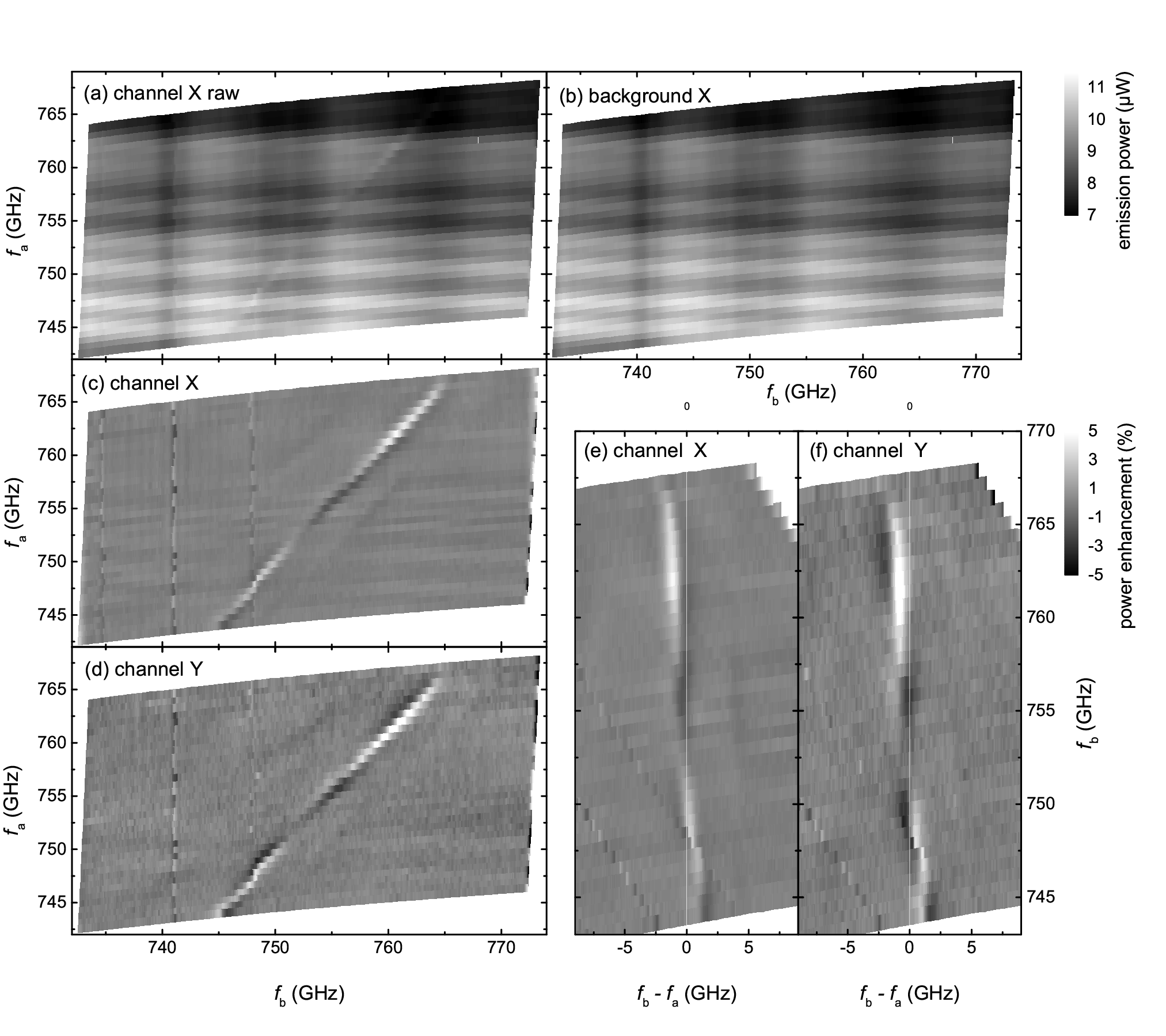}
	\caption{Emission power (gray scale) vs. frequencies $f_{\rm a}$ and  $f_{\rm b}$: (a): data as presented in Figure \ref{fig:tartanRawX}(b); (b): reconstructed tartan background; (c),(d): power enhancement $S_{\rm{\alpha}}(f_{\rm a},f_{\rm b})$ of the measured emission relative to the tartan-like background for graphs (a),(b) and (c) are for lock-in channel X, graph (d) is for lock-in channel Y.  Graphs (e) and (f) show the same data as graphs (c) and (b), but replotted in the ($f_{\rm b}-f_{\rm a}$,\,$f_{\rm a}$) plane. Vertical white lines in (e) and (f) indicate $f_{\rm b}-f_{\rm a}=0$.
	}
	\label{fig:tartanSubtracted}
\end{figure*}

In the next step we remove the tartan-like oscillations from the data in the ($f_{\rm a}$,  $f_{\rm b}$) plane. We start from the expression 
\begin{equation}
	\label{eq:totalPowerVoltage}
	\begin{split}
	P_{\rm{tot,\alpha}}(V_{\rm{r,a}},V_{\rm{r,b}}) &= P_{\rm{a,\alpha}}(V_{\rm{r,a}}) + P_{\rm{b,\alpha}}(V_{\rm{r,b}}) \\
	&+ P_{\rm{ab,\alpha}}(V_{\rm{r,a}},V_{\rm{r,b}}) .
	\end{split}
\end{equation}
We then extrapolate by an averaging procedure, which is outlined in Appendix \ref{Appendix B}, the tartan pattern over the diagonal line to find the ``background'' $B_{\rm{\alpha}}(V_{\rm{r,a}},V_{\rm{r,b}})$ which is given by the emission power of two independent stacks, 
\begin{equation}
	\label{eq:BackgroundVoltage}
		B_{\rm{\alpha}}(V_{\rm{r,a}},V_{\rm{r,b}}) = P_{\rm{a,\alpha}}(V_{\rm{r,a}}) + P_{\rm{b,\alpha}}(V_{\rm{r,b}}) .
\end{equation}
For clarity, Figure \ref{fig:tartanSubtracted}(a) repeats the measured data in the ($f_{\rm a}$,  $f_{\rm b}$) plane and Figure \ref{fig:tartanSubtracted}(b) shows the reconstructed tartan-like pattern for lock-in channel X. We then consider the ratio $S_{\rm{\alpha}}(f_{\rm a},f_{\rm b})=P_{\rm{tot,\alpha}}(f_{\rm a},f_{\rm b})/B_{\rm{\alpha}}(f_{\rm a},f_{\rm b})-1$, which yields the power enhancement ($S_{\rm{\alpha}}>0$) or attenuation ($S_{\rm{\alpha}}<0$) over the case of uncoupled IJJ stacks. Figures \ref{fig:tartanSubtracted}(c) and \ref{fig:tartanSubtracted}(d) show $S_{\rm{\alpha}}(f_{\rm a},f_{\rm b})$ for, respectively, the X channel and the Y channel of the lock-in. The ``diagonal'' line is the prominent feature in these plots and we find that $S_{\rm{\alpha}}(f_{\rm a},f_{\rm b})$ can be both positive and negative, with maximum values of about 0.05 in each direction. Further one notes in Figure \ref{fig:tartanSubtracted}(c) vertical lines near $f_{\rm b}=$\,735, 741 and 748\,GHz which have not been captured by the background reconstruction. In Figure \ref{fig:tartanSubtracted}(d) similar lines appear near 741 and 748\,GHz. These lines correspond to small but sudden changes in $P_{\rm{b,\alpha}}(f_{\rm b})$, which may have been caused by some IJJs in stack\;$b$ switching from the zero-voltage state to the resistive state. Finally, Figures. \ref{fig:tartanSubtracted}(e) and (f) respectively show $S_{\rm X}$ and $S_{\rm Y}$ as a function of $f_{\rm a}$ and the frequency difference $f_{\rm b}-f_{\rm a}$. In these plots the power enhancement/attenuation forms a nearly vertical line, with a small residual tilt which has not been captured by our correction procedure to obtain $V_{\rm{r,a}}$ and $V_{\rm{r,b}}$ and from there $f_{\rm a}$ and  $f_{\rm b}$.

Having found the frequency axes we can also assign the most likely cavity resonances that have been excited in the stacks. Generally, for a rectangular geometry the cavity resonance frequencies are given by $f_{\rm{cav}}=c_1\sqrt{(p/2W)^2+(q/2L)^2}$ , where $c_1\approx (6.5-7)\times 10^7$\,m/s is the in-phase mode velocity and the integers $p$ and $q$ count the number of half-waves of the $c-$axis electric field along, respectively, the width and the length of the stack. The emission occurs near 750\,GHz, which requires $p = 1$. For $q$ the most likely number is $3$. 

For the further analysis of $S_{\rm{\alpha}}(f_{\rm a},f_{\rm b})$, as shown in Figure. \ref{fig:tartanSubtracted}(e) and (f), let us first look at the presumed phase-synchronization of two IJJ stacks.  In an extremely simplified approach we start with an effective Kuramoto model \cite{Kuramoto75,Wiesenfeld96} of two coupled point-like identical oscillators. In the model, oscillator $a$ oscillates and radiates at a dressed (i.e., including interactions) angular frequency $\dot{\delta_{\rm a}} = \omega_{\rm a} + \frac{K}{2} \sin \delta$, and oscillator $b$ oscillates at $\dot{\delta_{\rm b}} = \omega_{\rm b} - \frac{K}{2} \sin \delta$. Here, $\delta=\delta_{\rm b}-\delta_{\rm a}$ is the phase difference between the two oscillators, $K$ is the strength of the mutual interaction. $\omega_{\rm{a}} = 2\pi \cdot f_{\rm{a}}$ and $\omega_{\rm{b}} = 2\pi \cdot f_{\rm{b}}$ are the bare frequencies without interaction. In case of phase-locking the dressed frequencies are equal, $\dot{\delta_{\rm a}}=\dot{\delta_{\rm b}}$ (common frequency $f_{\rm 0}$) and we obtain
\begin{equation}
	\label{eq:phaseDifference}
	\delta = \sin^{-1} \left( \frac{f_{\rm b}-f_{\rm a}}{K/2\pi} \right) .
\end{equation}
This constrains the phase difference to the interval $[-\pi/2, \pi/2]$ and implies $|f_{\rm b}-f_{\rm a}|\leq K/ 2\pi$ during phase-locking.  With respect to emission properties, ignoring optical path differences, these two oscillators would emit a power $P$ proportional to $|E_{\rm a}+E_{\rm b}e^{i\delta}|^2 = E_{\rm{0,a}}^2 + E_{\rm{0,b}}^2 + 2E_{\rm{0,a}}E_{\rm{0,b}} \cos(\delta)$ , where $E_{\rm a}=E_{\rm{0,a}}e^{i \theta_{\rm a}}$ and $E_{\rm b}=E_{\rm{0,b}}e^{i \theta_{\rm b}}$ denote the complex light fields. The expression may be reformulated as
\begin{equation}
	\label{eq:totalPowerCosine}
	P = P_{\rm a} + P_{\rm b} + 2\sqrt{P_{\rm a} P_{\rm b}} \cos(\delta) ,
\end{equation}
where $P_{\rm a}$ and $P_{\rm b}$ denote the power emitted by, respectively oscillator $a$ and oscillator $b$.  One notices, that when $\delta$ approaches $\pm \pi/2$ there is no difference to decoupled oscillators. For the signal enhancement $S$ one expects on this level
\begin{equation}
	\label{eq:S1}
	S = \frac{2\sqrt{P_{\rm a} P_{\rm b}}}{P_{\rm a} + P_{\rm b}} \cos(\delta) .
\end{equation}
Inserting Eq. (\ref{eq:phaseDifference}) one finds that a plot of $S$ vs. $f_{\rm b}-f_{\rm a}$ is just a semi-circle ending at $(f_{\rm b}-f_{\rm a})2\pi /K = \pm1$. When thermal fluctuations are included, phase-lock will not be granted over very long times. Particularly, we expect that the phase-lock goes to zero when the dressed frequencies are far from the degeneracy point $f_{\rm a}=f_{\rm b}$ of the undressed frequencies.  In Eq. (\ref{eq:S1}) $\cos(\delta)$ should be replaced by the time average $<\cos(\delta)>$ which we can formally also write as $<\cos(\delta)>=C(\Delta f)\cos(\delta_f)$ , with $C(\Delta f)=\sqrt{<\cos \delta>^2+<\sin \delta>^2}$, $\delta_f=\tan^{-1}(<\sin \delta>/<\cos \delta>)$ and $\Delta f = f_{\rm b}-f_{\rm a}$. Then, $S$ is given by 
\begin{equation}
	\label{eq:S2}
	S = \frac{2\sqrt{P_{\rm a} P_{\rm b}}}{P_{\rm a} + P_{\rm b}} C(\Delta f)\cos(\delta_f) .
\end{equation}

In a more realistic setting we have to consider that we have two stacks, each consisting of strongly coupled oscillators (the IJJs). In addition, a cavity resonance has built up in each stack causing spatial variations of the Josephson phases, and the presence of a hot spot complicates the situation even more. A single stack can be modelled numerically by using 3D coupled sine-Gordon equations combined with heat diffusion equations \cite{Rudau15,Rudau16}. In the code the $N$ IJJs in the stack are grouped to $M$ segments where the $G = N/M$ junctions in a given segment are assumed to oscillate coherently. To approach the situation of two coupled stacks to zero order we consider two electrically decoupled rectangular stacks separated by one pixel. Details are given in Appendix \ref{Appendix C}. The two stacks share a common continuous segment which can provide mutual coupling. The common segment can be fully superconducting representing a toy model for a base crystal or it can be normal conducting representing a toy model for the common gold electrode. Further, the thermal coupling between the stacks through the common segment and the substrate underneath remains intact. The basic result from this model is that the phase $\delta_f$ varies almost linearly between $\pm \pi$ as a function of $f_{\rm b}-f_{\rm a}$, and we see a decaying function  which, for simplicity we will approximate by a Gaussian for further analysis.  Expressed in terms of the frequency difference $f_{\rm b}-f_{\rm a}$ the correlation function reads:

\begin{equation}
	\label{eq:C}
	C(\Delta f)=C_{\rm 0}e^{-[(f_{\rm b}-f_{\rm a})/f_{\rm C}]^2} ,
\end{equation}
where $f_{\rm C}$ is a fit parameter. In the simulations shown in Appendix \ref{Appendix C} $C_{\rm 0}$ is of order 0.5. We further use the linear function, $\delta_f=(f_{\rm b}-f_{\rm a})/f_{\rm \delta}$,  where $f_{\rm \delta}$ is a fit parameter, to describe the functional dependence of $\delta_f$ on $f_{\rm b}-f_{\rm a}$. 

The final step, before the data shown in Figure \ref{fig:tartanSubtracted} can be analyzed further, is to find an expression analogous to Eq. (\ref{eq:S2}) for the actual lock-in detected functions $S_{\rm X}$ and $S_{\rm Y}$. Details are given in Appendix \ref{Appendix D}. The basic results are the expressions

\begin{subequations}
	\label{eq:S3}
	\begin{align}
	\begin{split}
	S_{\rm X} &= \frac{P_{\rm{X0}}}{P_{\rm{Xa}}+P_{\rm{Xb}}} C(\Delta f) \cos(\delta_f + \varphi_{\rm X}) \\
	&= S_{\rm{X0}} \, C(\Delta f) \cos(\delta_f + \varphi_{\rm X})\\
	\end{split}\\
	\begin{split}
	S_{\rm Y} &= \frac{P_{\rm{Y0}}}{P_{\rm{Ya}}+P_{\rm{Yb}}} C(\Delta f) \cos(\delta_f + \varphi_{\rm Y}) \\
	&= S_{\rm{Y0}} \, C(\Delta f) \cos(\delta_f + \varphi_{\rm Y}),\\
	\end{split}
	\end{align}
\end{subequations}
with $C(\Delta f)$ as in Eqs. (\ref{eq:S2}) and (\ref{eq:C}). Compared to Eq. (\ref{eq:S2}) there are extra phases $\varphi_{\rm X}$ and $\varphi_{\rm Y}$ which are different for the X and Y channels. They depend on parameters which cannot be extracted from experimentally available data and should be treated as free parameters. Due to the integration over the interfering beam profiles the amplitudes $S_{\rm{X0}}$ and $S_{\rm{Y0}}$ can go well below 1 and depend on unknown details, so that no direct conclusion on the magnitude $C_{\rm 0}$ of the correlation function $C(\Delta f)$ can be made.

\begin{figure}[h]
	\includegraphics[width=\columnwidth,clip]{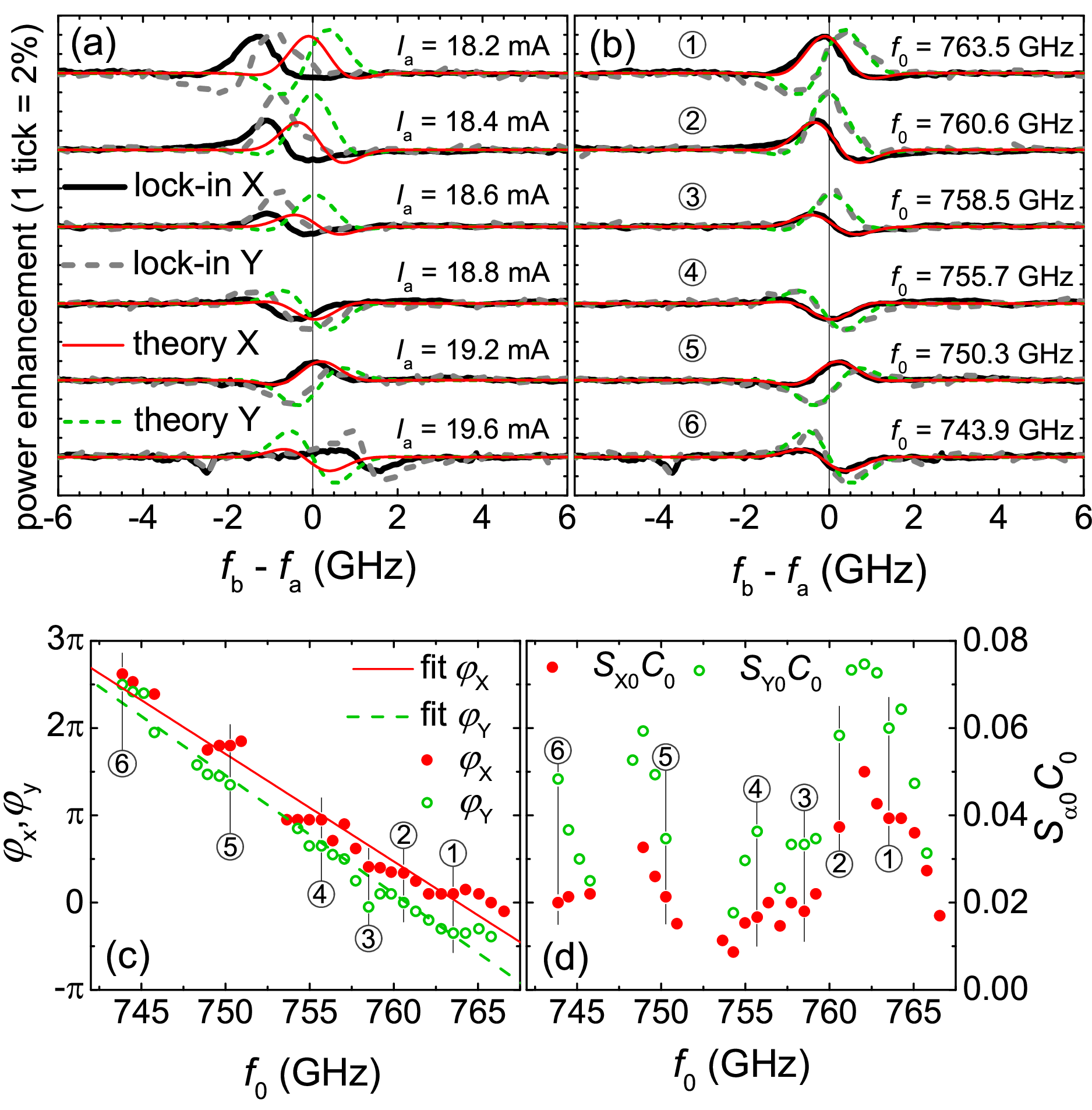}
	\caption{Evaluation of power enhancement. Graphs (a) and (b) show $S_{\rm \alpha}$ vs. frequency difference $f_{\rm b}-f_{\rm a}$ for 6 values of the common frequency $f_{\rm 0}$ where $f_{\rm b}=f_{\rm a}$. Currents $I_{\rm a}$ through stack\;$a$ are also indicated. Curves for different values of $f_{\rm 0}$ are vertically offset for clarity. Experimental data for, respectively, the X and Y channel are given by the solid black and the dashed grey lines. Solid red and dashed green lines represent corresponding theoretical curves using Eqs. (\ref{eq:S3}a) and (\ref{eq:S3}b). Model parameters are chosen to best reproduce the shape of the experimental curves. Graph (a) shows the original data. In graph (b) the experimental curves are shifted to be on top of the theoretical curves. Graph (c) displays the phases $\varphi_{\rm X}$ and $\varphi_{\rm Y}$ vs. $f_{\rm 0}$. Linear fits are shown by the solid lines. In Graph (d) the factor $S_{\rm{\alpha 0}}C_{\rm 0}$ required to match the amplitudes of the experimental and theoretical curves of $S_{\rm \alpha}$ is displayed vs. the common frequency $f_{\rm 0}$. Labels \raisebox{.5pt}{\textcircled{\raisebox{-.9pt} {1}}} to \raisebox{.5pt}{\textcircled{\raisebox{-.9pt} {6}}} in (c), (d) indicate the linescans \raisebox{.5pt}{\textcircled{\raisebox{-.9pt} {1}}} to \raisebox{.5pt}{\textcircled{\raisebox{-.9pt} {6}}} labelled in (b).  
	}
	\label{fig:linescans}
\end{figure}

Figure \ref{fig:linescans}(a) shows by the black solid (X channel) and grey dashed (Y channel) lines experimental curves of $S_{\rm \alpha}$ vs. $f_{\rm b}-f_{\rm a}$ for 6 values of the common frequency $f_{\rm 0}$ where $f_{\rm b}=f_{\rm a}$.  $S_{\rm \alpha}$ shows a variation of maximally 5\% when scanning the difference frequency $f_{\rm b}-f_{\rm a}$. For technical reasons (comparatively large current steps between adjacent values of the bias current $I_{\rm a}$) we used the original curves that were recorded at a fixed bias current $I_{\rm a}$. I.e., compared to the plots in Figures \ref{fig:tartanSubtracted}(e) and (f) the line scans are slightly tilted to the horizontal direction, which, however, makes little difference to the data evaluation in the narrow frequency difference regime we display in Figure \ref{fig:linescans}(a). The currents $I_{\rm a}$ are indicated. The corresponding common frequencies $f_{\rm 0}$ are indicated in Figure \ref{fig:linescans}(b). The six curves are vertically offset for clarity. Note that in the spirit of phase-locking, as described above, the horizontal axis should display the difference of the bare frequencies (decoupled oscillators) rather than the measured difference of the dressed frequencies (coupled oscillators). However, this difference is unresolvable in the presented experiments and we thus do not distinguish them here. Further, the experimental data are not centered around $f_{\rm b}-f_{\rm a}=0$. We attribute this to residual errors in our evaluation procedure to extract $f_{\rm a}$ and $f_{\rm b}$. To compare the experimental data with the theoretical expressions of Eqs. \ref{eq:S3}(a) and \ref{eq:S3}(b) we first find parameters that optimally reproduce the shape of $S_{\rm \alpha}$ vs. $f_{\rm b}-f_{\rm a}$. We first globally set $f_{\rm C}=$\;0.9\,GHz and $f_{\rm \delta}=$\;0.58\,GHz for all curves. We then adjust the phases $\varphi_{\rm X}$ and $\varphi_{\rm Y}$ for each curve and individually adjust the factor $S_{\rm{\alpha 0}}C_{\rm 0}$ to match the amplitude of the theoretical curves and the experimental ones. The results are shown by the solid red and dashed green lines representing, respectively, the X and Y channel. As dictated by Eqs. \ref{eq:S3}(a) and \ref{eq:S3}(b) the theoretical curves are centered around $f_{\rm b}-f_{\rm a}=0$.  For a better comparison Figure \ref{fig:linescans}(b) shows the same data but with the experimental curves shifted to be on top of the theoretical curves. The agreement is excellent, making us confident that the essential physics is captured by Eqs. \ref{eq:S3}(a) and \ref{eq:S3}(b).

Figure \ref{fig:linescans}(c) displays the phases $\varphi_{\rm X}$ and $\varphi_{\rm Y}$ vs. the common frequency $f_{\rm 0}$. Apart from the constant offset the basic observation is that both $\varphi_{\rm X}$ and $\varphi_{\rm Y}$ are, within our resolution, linear functions with slopes $\mathrm{d} \varphi_{\rm{\alpha}} / \mathrm{d} f_{\rm 0}$ of, respectively $-2\pi/16.2$\,GHz and $-2\pi/14.8$\,GHz for the X and the Y channel. The linear dependence of $\varphi_{\rm X}$ and $\varphi_{\rm Y}$ on $f_{\rm 0}$ is in-line with our assumption that between the location of the oscillators and the location of the bolometer substantial geometric phase differences (of unknown origin) have built up. Interpreted in terms of optical path differences between the two stacks the observed slopes would imply an \textit{average} path difference $\Delta l$ of 2\,cm, which is still too large to be explained easily.  Finally, The amplitudes $S_{\rm{\alpha 0}}C_{\rm 0}$ are displayed in Figure \ref{fig:linescans}(d). This amplitude can exceed 7\% but, near 755\,GHz can also be below 2\%. The overall dependence of $S_{\rm{\alpha 0}}C_{\rm 0}$ on $f_{\rm 0}$ is not very systematic and may have been caused by small drifts and fluctuations on the time scale of several minutes or so in our detection scheme. While the effect for the overall detected power is small, such fluctuations are strongly enhanced when calculating $S_{\rm{\alpha 0}}C_{\rm 0}$. More importantly, the overall small values of $S_{\rm{\alpha 0}}C_{\rm 0}$ are consistent with our assumption that the wave fronts created by the two stacks show very strong spatial phase gradients, leading to destructive interference in the detection plane. 

We finally mention that the numerical simulations outlined in Appendix \ref{Appendix C} delivered identical time-averaged voltages and Josephson frequencies between the two stacks only within a part (about less than half) of the frequency interval $f_{\rm C}$ spanned by the Gaussian. The true locking range $|f_{\rm b}-f_{\rm a}|$ may thus be below 200\,MHz, which is consistent with our previous measurements with the SIR, where two individual lines appeared when their emission frequencies differed by around 100\,MHz.

\section{Summary and Conclusions}
\label{sec:Summary}

As our main result we observed mutual synchronization of two stacks of IJJs in a standalone geometry without common base crystal. The distance between the stacks is 200\,$\mu$m and they share a common gold electrode presumably providing mutual coupling. Some preliminary data were obtained within a high spectral resolution setup utilizing a superconducting integrated receiver at around 650\,GHz \cite{Koshelets19}. The measurements presented in the present manuscript are based on a bolometric emission setup and extend the observation of phase-lock for the same sample to frequencies between 745 and 765\,GHz. Since the present setup is a standard one available in many laboratories our measurement protocol and the subsequent detailed analysis may be helpful for many other groups working in the field.

We find that phase correlations between the two stacks have Gaussian shape and occur on the scale of about $\pm0.4$\,GHz relative to the center frequencies where the bare (uncoupled) frequencies of the two oscillators coincide. We give arguments that the wave fronts emitted by the two stacks have very strong spatial gradients, leading to destructive interference effects. We speculate that the origin lies in the excited cavity modes which, on the one hand, seem to be necessary for synchronization but, on the other hand, lead to a strongly varying Josephson phase along the edges of each stack.

Our observation, that the mutual synchronization of IJJ stacks separated by relatively large distances can be induced by a common gold electrode rather than a base crystal, could relax the notorious problem of Joule heating, because the gold layer can provide a much better thermal coupling to a substrate than the BSCCO base crystal. In addition, the problem of strong spatial phase gradients may be relaxed, if the individual stacks in a multi-stack array are made considerably smaller in lateral size than the presently used ones. Thus, the gold-layer-mediated coupling may open new possibilities to synchronize a large number of IJJs to increase the emission power to values well above 1\,mW.

\acknowledgments
We gratefully acknowledge financial support by the Deutsche Forschungsgemeinschaft via projects KL930/13-2 and KL930/17-1,
The Scientific and Technological Research Council of T\"urkiye, T\"UB\.ITAK 2219-2020/2-1059B192000817, the Research Fund of Inonu University, T\"urkiye under Grant Contract No. FOA-2023-3260, the Russian Science Foundation, grant No. 19-19-00618. the National Key R\&D Program of China, grant No. 2021YFA0718802, the National Natural Science Foundation of China, grant No. 62288101, and Jiangsu Key Laboratory of Advanced Techniques for Manipulating Electromagntic Waves. R. Wieland acknowledges financial support from the Carl Zeiss Stiftung.

\appendix
\renewcommand\thefigure{\thesection.\arabic{figure}}

\section{Conversion from current axes to voltage and frequency axes, determination of contact resistances and junction numbers }
\setcounter{figure}{0}
\label{Appendix A}

The voltages $V_{\rm a}$ and $V_{\rm b}$ which we measure contain parasitic contributions from the contact resistances and perhaps from IJJs which do not contribute to the (coherent) radiation of the single stacks. We write these contributions as $V_{\rm{p,a}} = R_{\rm a}I_{\rm a}$ for stack\;$a$ and  as $V_{\rm{p,b}} = R_{\rm b}I_{\rm b}$ for stack\;$b$.  We thus have  $V_{\rm{r,a}} + R_{\rm a}I_{\rm a} = V_{\rm a}$ and $V_{\rm{r,b}} + R_{\rm b}I_{\rm b} = V_{\rm b}$. $V_{\rm{r,a}}$ and $V_{\rm{r,b}}$ are the voltages of interest. In general, $R_{\rm a}$ and $R_{\rm b}$ will be nonlinear with respect to the bias currents and may, because of heating, also depend on the biasing condition of the other stack. To determine them we can make use of previous observations that the quasiperiodic oscillations in the detected emission power as a function of either current or voltage, c.f. Figures. \ref{fig:IVC}(b) and \ref{fig:IVC}(c), are actually functions of the frequencies of the emitted radiation. This means, in a plot of the detected radiation power vs. $V_{\rm{r,a}}$ and $V_{\rm{r,b}}$ they should occur as horizontal and vertical stripes. Actual values for $R_{\rm a}$ and $R_{\rm b}$ that fulfill this condition to a good accuracy within the current range $17.9\,\mathrm{mA} < I_{\rm a} < 19.7\,\mathrm{mA}$ and $20\,\mathrm{mA} < I_{\rm b} < 25\,\mathrm{mA}$ and used to generate Figure \ref{fig:tartanRawX}(b) are $R_{\rm a}=0.5\,\Omega-4 \Delta I_{\rm b}\Omega/\rm{A}$ and $R_{\rm b}=2.8\,\Omega$, where $\Delta I_{\rm b}=I_{\rm b}-20$\,mA. With current for stack\;$b$ maximally varying by $\Delta I_{\rm b} = 5$\,mA, $R_{\rm a}$ varies in the range from 0.48\,$\Omega$  to 0.5\,$\Omega$. 

To convert voltages $V_{\rm{r,a}}$ and $V_{\rm{r,b}}$ to emission frequencies, the junction numbers $N_{\rm a}$ and $N_{\rm b}$ must be known. To find these numbers, in a first step we measure the emission frequencies using our Fourier-transform interferometer and relate them to voltages $V_{\rm{r,a}}$ and $V_{\rm{r,b}}$. In a second step we refine the calibration by making use of the water vapor absorption line at $f_{\mathrm{water}} \approx$\,752.033\,GHz which becomes visible in absorption spectra when the interferometer is flooded with (humid) air rather than with nitrogen gas. 

\begin{figure}[h]
	\includegraphics[width=\columnwidth,clip]{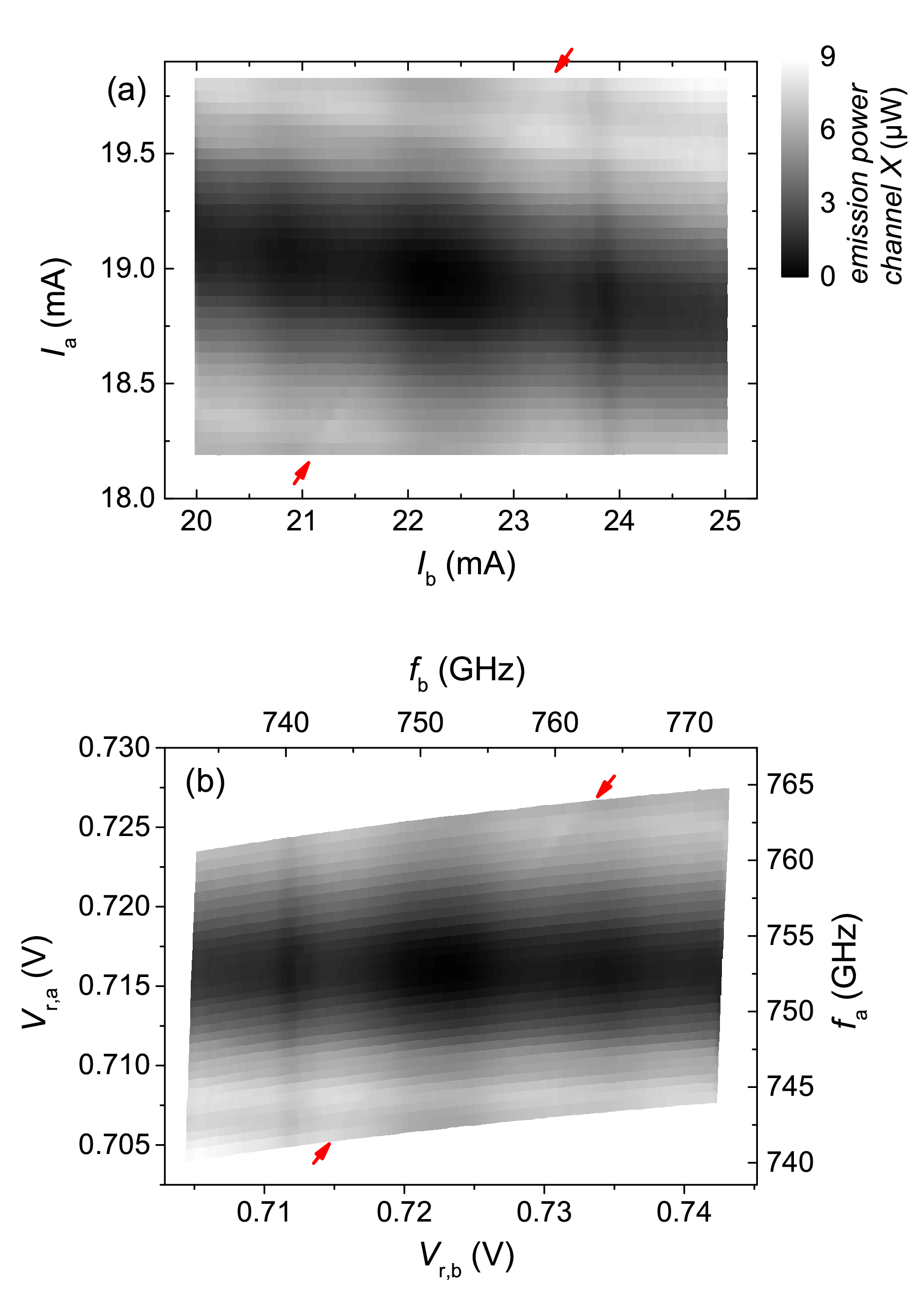}
	\caption{For lock-in channel X: Emission data for the two stacks biased at the same time, while the setup was exposed to humid air, to be compared with Figure \ref{fig:tartanRawX} of the main text. (a) Bolometric power as grayscale plotted against stack currents $I_{\rm a}$ and $I_{\rm b}$. (b) Bolometric data as a function of the rescaled voltages $V_{\rm{r,a}}$ and $V_{\rm{r,b}}$ (left and bottom axes) and the corresponding emission frequencies $f_{\rm a}$ and $f_{\rm b}$  created from the voltage axes using $N_{\rm a}=$\,460 and $N_{\rm b}=$\,465. (right and top axes).  
	}
	\label{fig:tartanRawXair}
\end{figure}

\begin{figure}[h]
	\includegraphics[width=\columnwidth,clip]{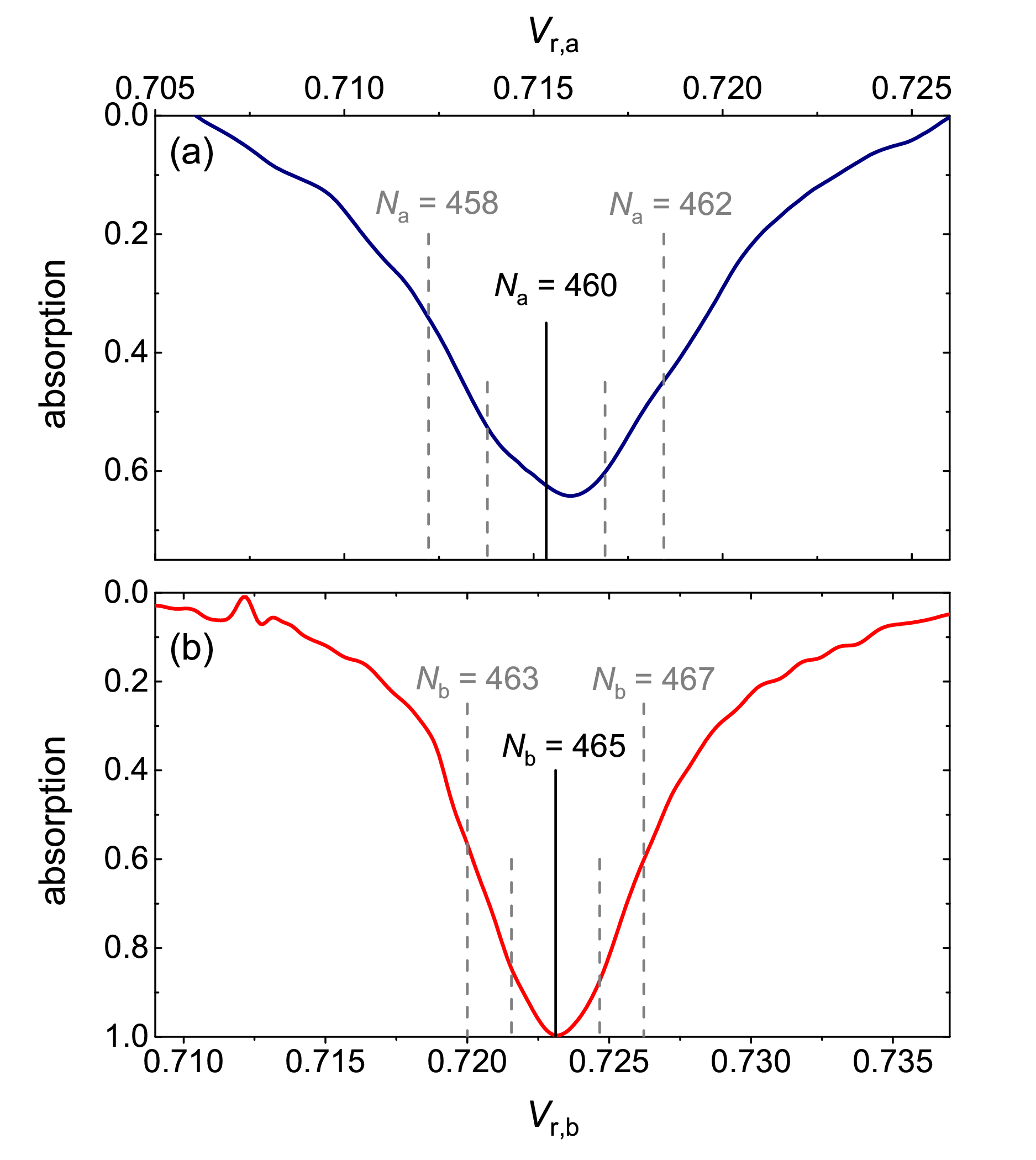}
	\caption{Absorption spectra as functions of rescaled stack voltages for stack\;$a$ (a) and stack\;$b$ (b). Voltages corresponding to different junction numbers are indicated by vertical lines, the solid ones indicating the most likely numbers.
	}
	\label{fig:absorption}
\end{figure}

To do so we have both oscillators in the emitting state, as shown in Figure \ref{fig:tartanRawX}, where the bias currents through the two stacks are varied over some range while the setup was flooded with nitrogen. We then perform the same measurement with the setup exposed to humid air. The result is shown in Figure \ref{fig:tartanRawXair}(a) in the ($I_{\rm a},I_{\rm b}$) plane. Figure \ref{fig:tartanRawXair}(b) shows the same data in the ($V_{\rm{r,a}}, V_{\rm{r,b}}$) plane. As outlined in Appendix \ref{Appendix B} the data of Figure \ref{fig:tartanRawX} and Figure \ref{fig:tartanRawXair} allow a reconstruction of the THz power emitted by the individual stacks as a function of, respectively, $V_{\rm{r,a}}$ and $V_{\rm{r,b}}$.
Using these curves we create the normalized difference 
$(P_{\rm{k,N_2}}-P_{\rm{k,air}})/(P_{\rm{k,N_2}}+P_{\rm{k,air}})$, with $k = (a,b)$. This difference constitutes the absorption and is plotted in Figure \ref{fig:absorption} for the two stacks vs., respectively, voltages $V_{\rm{r,a}}$ and $V_{\rm{r,b}}$. Graph \ref{fig:absorption}(a) is for stack\;$a$ and graph \ref{fig:absorption}(b) for stack\;$b$. The absorption line for stack\;$a$ leads to $N_{\rm a}=$\,460$\pm$1. 
For  stack\;$b$, $N_{\rm b}=$\,465 marks the most likely junction number.  Using these values we create the frequency axes shown in Figure \ref{fig:tartanRawX}(b), Figure \ref{fig:tartanSubtracted} and in Figure \ref{fig:tartanRawXair}(b).

\section{Reconstruction of the emission background created by unlocked stacks}
\setcounter{figure}{0}
\label{Appendix B}

To find the tartan-like background of the recorded emission power in the ($V_{\rm{r,a}}, V_{\rm{r,b}}$) plane we start from expression (\ref{eq:totalPowerVoltage}) of the main paper,
\begin{equation}
	\label{eq:totalPowerVoltageAppendixB}
	\tag{\ref{eq:totalPowerVoltage}}
	\begin{split}
		P_{\rm{tot,\alpha}}(V_{\rm{r,a}},V_{\rm{r,b}}) &= P_{\rm{a,\alpha}}(V_{\rm{r,a}}) + P_{\rm{b,\alpha}}(V_{\rm{r,b}}) \\
		&+ P_{\rm{ab,\alpha}}(V_{\rm{r,a}},V_{\rm{r,b}}),
	\end{split}
\end{equation}
where $B_{\rm{\alpha}}(V_{\rm{r,a}},V_{\rm{r,b}}) = P_{\rm{a,\alpha}}(V_{\rm{r,a}}) + P_{\rm{b,\alpha}}(V_{\rm{r,b}})$ forms the background signal created by two independent IJJ stacks and $\alpha$ stands for either the X channel or the Y  channel. As we know from emission measurements of the stacks being biased one at a time we know that $P_{\rm{a,\alpha}}(V_{\rm{r,a}})$ and $P_{\rm{b,\alpha}}(V_{\rm{r,b}})$ exhibit modulations as a function of, respectively, $V_{\rm{r,a}}$ and $V_{\rm{r,b}}$. Further, these oscillations appear on top of a constant background which, when both stacks are biased simultaneously, cannot be unambiguously attributed to either stack\;$a$ or stack\;$b$. For simplicity we add this contribution to stack\;$a$ in the fit-procedure.

We thus, independently for the X and Y channels, parameterize $P_{\rm{a,\alpha}}(V_{\rm{r,a}})$ as the sum of several Gaussian curves plus a linear term 
\begin{equation}
	\label{eq:fitPa}
	\begin{split}
		P_{\rm{a,\alpha}}(V_{\rm{r,a}}) &= b + m \cdot V_{\rm{r,a}} \\
		&+ \sum_{n=1}^{N_A} a_{n,\rm{a}} \exp \left[-\left[\frac{V_{\rm{r,a}}-v_{n,\rm{a}}}{c_{n,\rm{a}}}\right]^2\right]
	\end{split}
\end{equation}
with $N_A=22$.

For $P_{\rm{b,\alpha}}(V_{\rm{r,b}})$ we use
\begin{equation}
	\label{eq:fitPa}
		P_{\rm{b,\alpha}}(V_{\rm{r,b}}) = \sum_{n=1}^{N_B} a_{n,\rm{b}} \exp \left[-\left[\frac{V_{\rm{r,b}}-v_{n,\rm{b}}}{c_{n,\rm{b}}}\right]^2\right]
\end{equation}
with $N_B=24$.

\begin{figure}[h]
	\includegraphics[width=\columnwidth,clip]{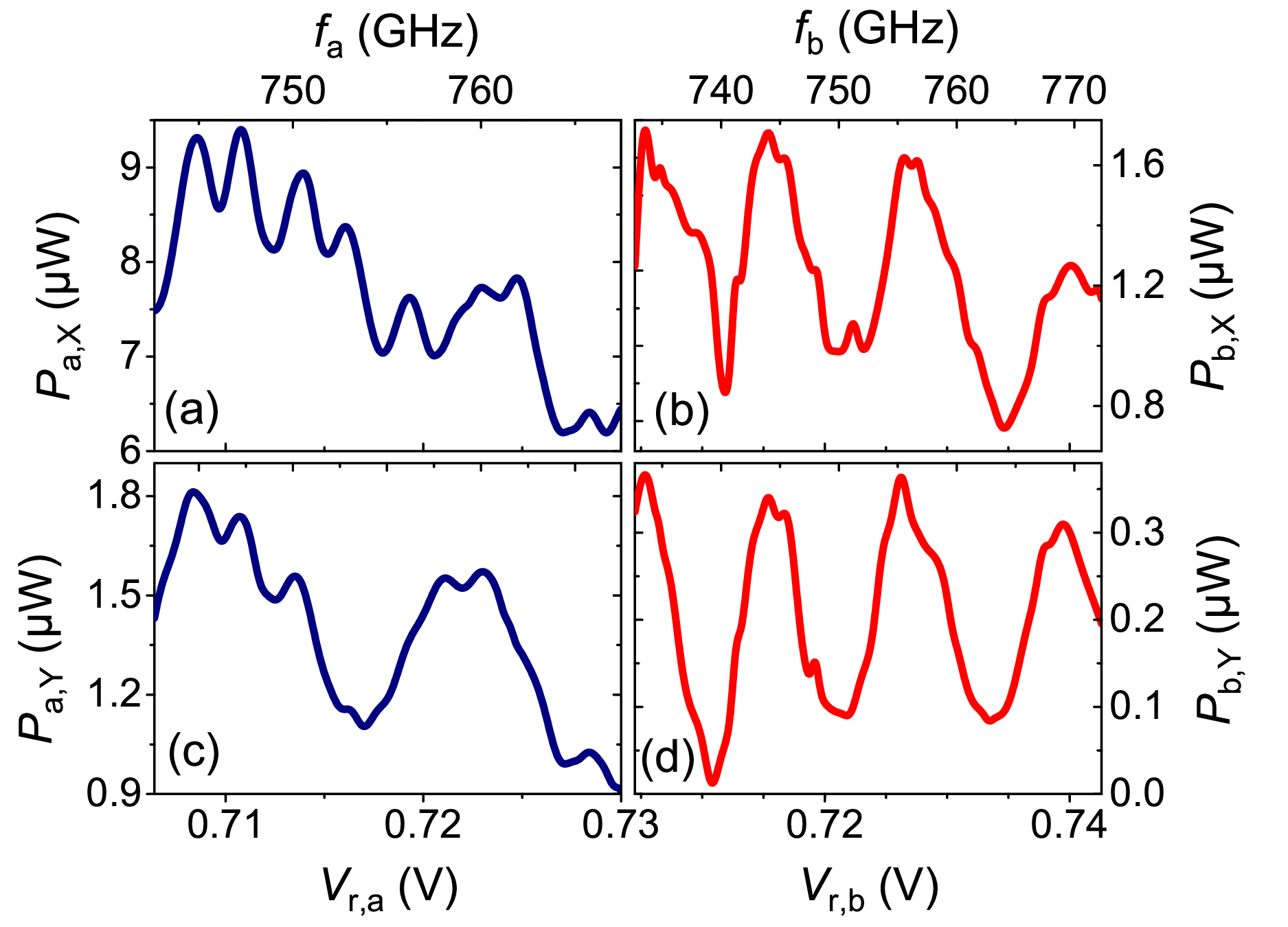}
	\caption{Functions $P_{\rm{a,X}}$ (a), $P_{\rm{b,X}}$ (b), $P_{\rm{a,Y}}$ (c) and $P_{\rm{b,Y}}$ (d), as obtained from the fit procedure.
	}
	\label{fig:fitResults2x2}
\end{figure}

The resulting functions $P_{\rm{a,X}}(V_{\rm{r,a}})$ and $P_{\rm{b,X}}(V_{\rm{r,b}})$ with best fitting parameters are plotted in Figures. \ref{fig:fitResults2x2}(a) and  \ref{fig:fitResults2x2}(b) respectively. The background $B_X(V_{\rm{r,a}},V_{\rm{r,b}})$ evaluated at the voltages of all data points is plotted in the main article as Figure \ref{fig:tartanSubtracted}(b) with voltages converted to frequencies.  Figures. \ref{fig:fitResults2x2}(c) and (d) show the corresponding functions $P_{\rm{a,Y}}(V_{\rm{r,a}})$ and $P_{\rm{b,Y}}(V_{\rm{r,b}})$ for the Y channel.

As mentioned we cannot assign the constant $b$ unambiguously to stacks\;$a$ and $b$. A redistribution can be made from the analysis of the humid-air absorption data discussed in Appendix \ref{Appendix A}. By simply demanding that the absorption should not be negative we find that for the X cannel a constant offset of 0.625\;$\mu$W should be subtracted from stack\;$a$ and added to stack\;$b$. For the Y channel the corresponding number is 0.025\;$\mu$W. Necessary corrections strongly depend on actual fit results.

\begin{figure}[h]
	\includegraphics[width=\columnwidth,clip]{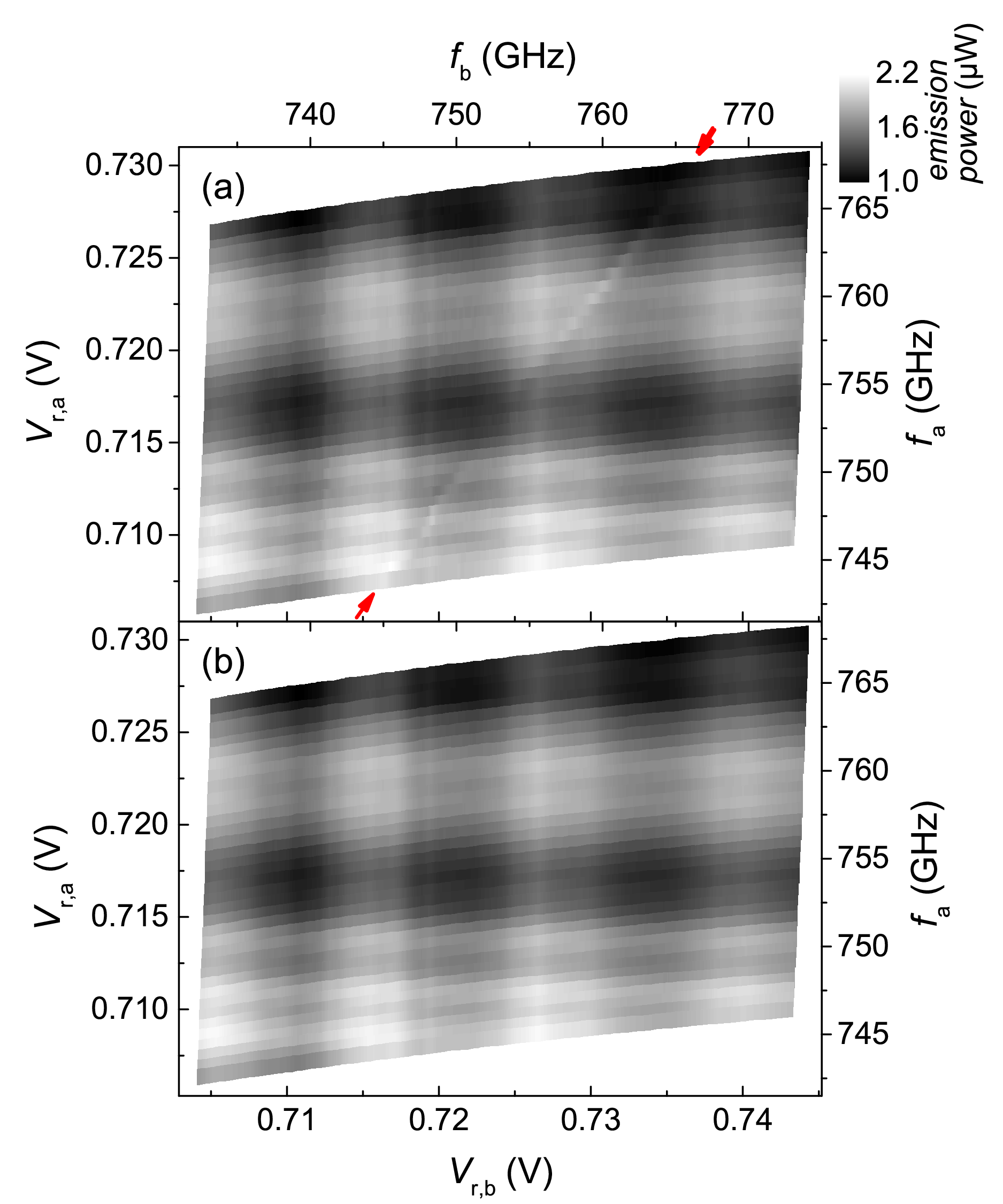}
	\caption{(a) raw emission data for channel Y. (b) reconstructed tartan-like background for channel Y.
	}
	\label{fig:tartanRawY}
\end{figure}

For completeness, Figure \ref{fig:tartanRawY}(a) shows the original emission data for channel Y in the ($V_{\rm{r,a}}, V_{\rm{r,b}}$) plane, while Figure \ref{fig:tartanRawY}(b) shows the reconstructed background $B_Y(V_{\rm{r,a}}, V_{\rm{r,b}})$.

We finally note that we have also 2D adjacent averaging methods to reconstruct the background. The results for the constituting functions and the reconstructed background are very similar. We are thus confident that the reconstruction is robust and does not suffer from major artifacts. 

\section{Minimal model for mutual phase-lock and simulation results }
\setcounter{figure}{0}
\label{Appendix C}
In the following we want so substantiate the expression
\begin{equation}
	\label{eq:S2AppendixB}
	\tag{\ref{eq:S2}}
	S = \frac{2\sqrt{P_{\rm a} P_{\rm b}}}{P_{\rm a} + P_{\rm b}} C(\Delta f)\cos(\delta_f)
\end{equation}
used in the main part of the paper, with a focus on the Gaussian approximation for $C(\Delta f)$, plus the linear approximation $\delta_f=(f_{\rm b}-f_{\rm a})/f_{\rm \delta}$.

To simulate two coupled stacks we start from 3D coupled sine-Gordon equations combined with heat diffusion equations \cite{Rudau15,Rudau16}. In the code we use, the $N$ IJJs in the stack are grouped to $M$ segments, where the $G = N/M$ junctions in a given segment are assumed to oscillate coherently. The IJJs are stacked in $z$ direction and extend over distances $L$ and $W$ in $(x,y)$ direction. 
For a single stack the electromagnetic part of the equation reads

\begin{equation}
	\label{eq:sigo_segment}
	\tag{C1a}
	\begin{split}
		Gsd\vec{\nabla}\left(\frac{\vec{\nabla}\dot{\gamma}_n}{\rho_{ab}(T)}\right) + G\lambda_{\rm k}^2 \vec{\nabla}\left[n_s(T)\cdot\vec{\nabla}\gamma_n\right] = \\
		\left(2+\frac{G^2\lambda_{\rm k}^2n_s(T)}{\lambda_{\rm c}^2}\right)j_{z,n}-j_{z,n+1}-j_{z,n-1}.
	\end{split}
\end{equation}
with
\begin{equation}
	\label{eq:RCSJ}
	\tag{C1b}
	j_{{z},n} = \frac{\beta_{\rm c}}{G} \ddot{\gamma}_n + \frac{\dot{\gamma}_n}{\rho_c(T)} + j^{\rm N}_{{z},n} + j_{\rm{c},n}(T) \sin(\gamma_n).
\end{equation}
The Nabla operators stand for partial derivative with respect to the $x$ and $y$ coordinates and $n$ counts the segments. $\gamma_n$ is the Josephson phase difference across a single junction in segment $n$. $s = 1.5$\,nm and $d = 0.3$\,nm are, respectively the interlayer distance and the thickness of a superconducting layer,  $\lambda_{\rm k} \approx 1.7\,\mu$m is the kinetic length,  $\lambda_{\rm c} \approx 300\,\mu$m is the $c$-axis penetration depth. $n_s$ is the Cooper pair density, normalized to its 4.2\,K value. $\rho_{ab}$ and $\rho_c$ denote the normalized in-plane and $c$-axis resistivities, and $\beta_{\rm c}$ is the McCumber parameter. The first term on the left-hand side of Eq. (\ref{eq:sigo_segment}) represents the (inductive) coupling via in-plane quasiparticle currents, while the second term is due to the coupling via supercurrents.

Equations (\ref{eq:sigo_segment}) and (\ref{eq:RCSJ}) form the coupled sine Gordon equations, with the $c$-axis currents given by Eq. (\ref{eq:RCSJ}). Here, the terms on the right-hand side represent the densities of displacement currents, quasiparticle currents, a noise current and the Josephson current. Various parameters depend on the local temperature $T$ as indicated. The Joule heat power density $q$ produced by the resistive in-plane and out-of-plane currents, plus the Joule heat generated by currents through a bond wire attached to the uppermost layer of the stack enters the heat-diffusion part of the equations,
\setcounter{equation}{1} 
\begin{equation}
	\label{eq:heat}
	c \dot{T} = \vec{\nabla} \left[ \kappa(T) \vec{\nabla}T \right] + q(T),
\end{equation}
where $\kappa$ is the (anisotropic) heat conductivity. Equations (C1) 
 and (\ref{eq:heat}) are solved simultaneously.

To obtain a certain value of, say, the dc voltage across the stack we first perform an initialization sequence in the absence of Josephson currents to get an initial guess of the temperature distribution in the stack. We then propagate equations (C1) and (\ref{eq:heat}) in time over about 20000 periods of the Josephson oscillations to relax the system and then take time traces over another 20000 periods to obtain time averages and other quantities. More details are given in \cite{Rudau15,Rudau16}.

\begin{figure}[h]
	\includegraphics[width=\columnwidth,clip]{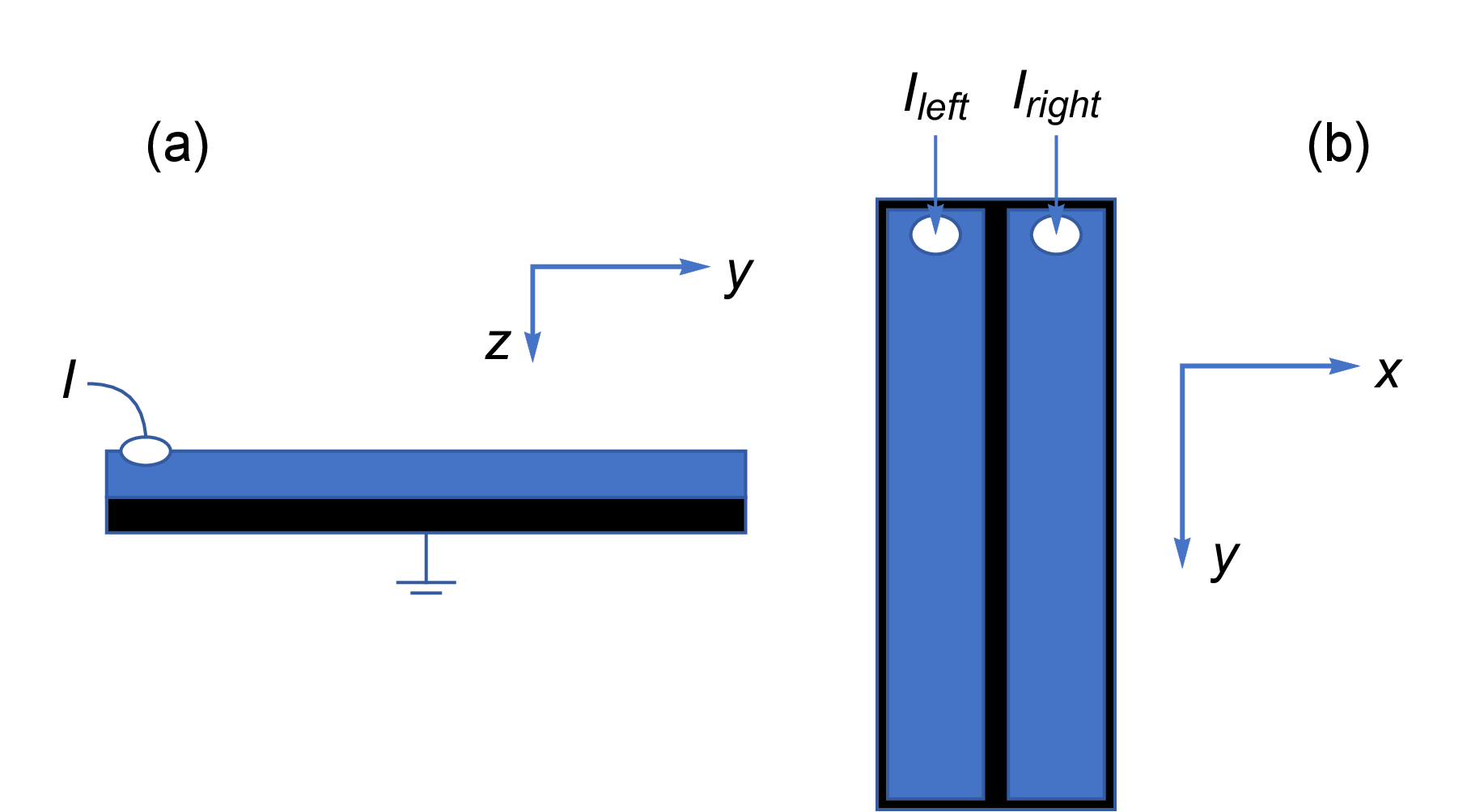}
	\caption{Geometry of two stacks located on a common segment. (a) side view. (b) top view. In (a) the black rectangle represents the non-cut base segment.
	}
	\label{fig:geometryCSG}
\end{figure}

To describe two coupled stacks we start from a solid stack of length $L$ and width $2W$ and consider $M+1$ segments. The geometry is sketched in Figure \ref{fig:geometryCSG}. We remove the coupling mediated by the terms on the left-hand side of Eq. (\ref{eq:sigo_segment}) in the upper $M$ segments along a cut extending in $y$ direction. I.e., we do not allow supercurrents and resistive currents to flow from the right side of the left stack to the left side of the right stack. We leave the lowest segment intact. This constitutes two decoupled $M$ segment stacks located on a common base segment, the lower surface of which we assume to be grounded. 

When we keep the electrical parameters as given in Eq. (C1) we have a minimal model for two side-by-side stacks located on a common base crystal. Further, to mimic a normal conducting segment, we interrupt the supercurrent flow between the two stacks by setting also for the base segment the second term in the left-hand side of Eq. (\ref{eq:sigo_segment}) to zero and null the Josephson current by setting the current density $j_{\rm c}(T)$ to zero. For a gold layer one should ideally replace $\rho_{ab}$ and $\rho_c$ by the gold resistivities. However, it turned out that particularly a change in $\rho_c$ by orders of magnitude results in unacceptable computing times. We thus just reduced $\rho_{ab}$ and $\rho_c$ by a factor of 100 in the simulations shown below.

Note that in our toy model the two stacks remain fully coupled thermally. Also, their virtual distance is infinitely small. 

In the simulations shown below we describe each stack only by a single pixel in $x$ direction. In other words, we model two coupled 2D stacks. For the discretization in $y$ direction we use 50 points. The (independent) $M = 20$ segments contain $N = 700$ IJJs, i.e., $G = 35$. This constitutes a base segment with a thickness of 35 layers, i.e., 50\,nm. The bath temperature for the calculations is 50\,K, and we use $L = 300\, \mu$m and $W = 50\, \mu$m. These parameters do not exactly match the experiment. Particularly, a bath temperature of 50\,K has been chosen to obtain stable low-bias and high-bias regimes involving cavity resonances just by changing the bias current.

\begin{figure}[h]
	\includegraphics[width=\columnwidth,clip]{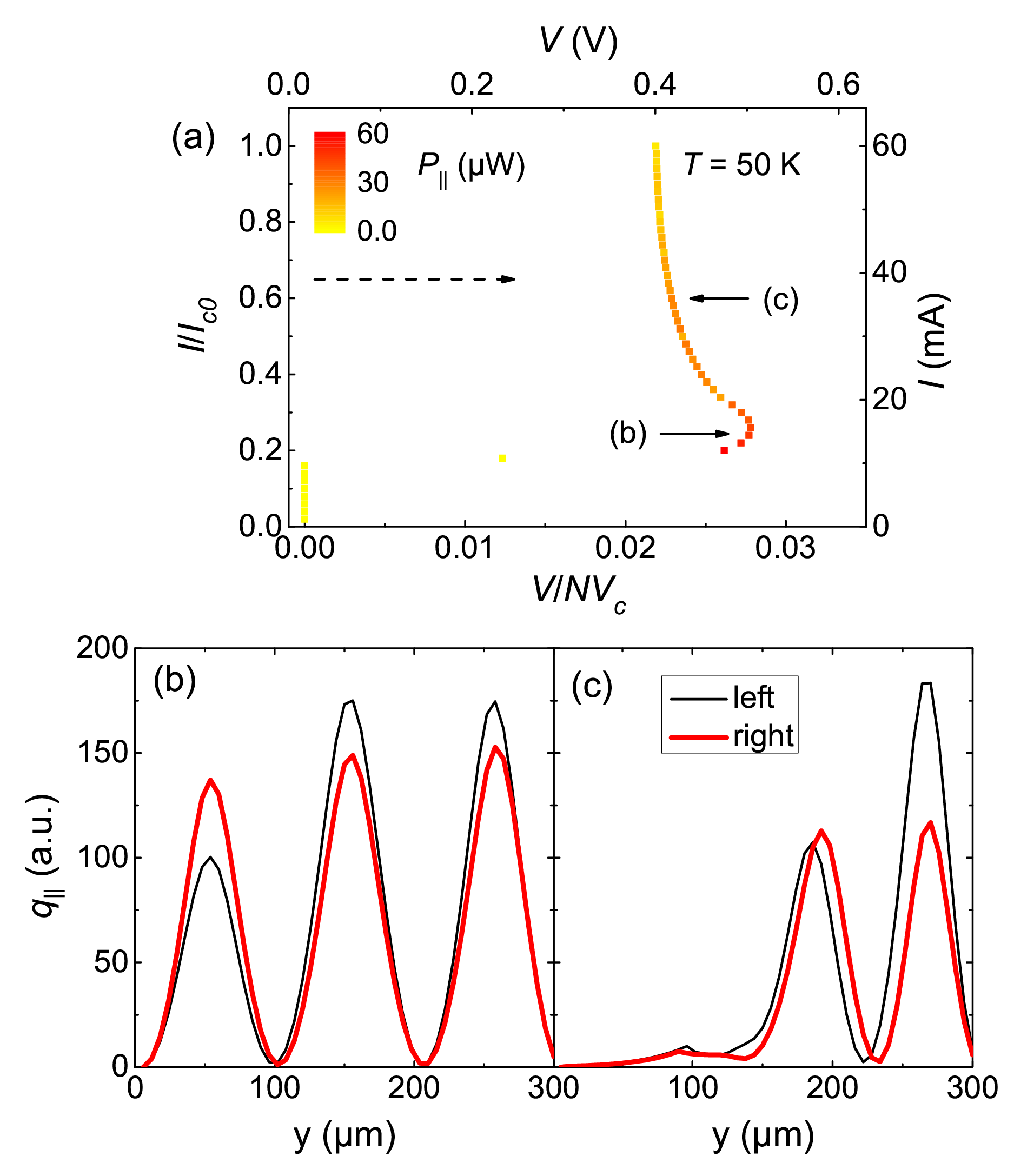}
	\caption{(a) Simulated current voltage characteristic of the uncut 300 $\times$ 100\,$\mu \rm m^2$ wide stack at a bath temperature of 50\,K. The color scale indicates the power dissipated by in-plane currents. The horizontal dashed arrow indicates switching from the zero-voltage state to the resistive state at the 50 K critical current. The horizontal solid arrows indicate the bias points used for biasing the left stack after cutting. (b) and (c) show simulation results for the two 300 $\times$ 50\,$\mu \rm m^2$ wide stacks are coupled via a normal conducting base segment. (b) shows the dissipated power density of the resistive in-plane currents for a bias $i_{\rm{left}} = 0.24$ and $i_{\rm{right}} = 0.235$. (c) shows the dissipated power density of the resistive in-plane currents for a bias $i_{\rm{left}} = 0.6$ and $i_{\rm{right}} = 0.58$.
	}
	\label{fig:IVCcsg}
\end{figure}

Figure \ref{fig:IVCcsg} shows some results to characterize the stacks investigated. Figure \ref{fig:IVCcsg}(a) displays the 50\,K current voltage characteristic (IVC) of the $L\times 2W$ wide stack before cutting. On the left vertical axis the current is given in units of the 4.2\,K value $I_{\rm{c} 0} = 60$\,mA of the critical current while the right axis displays the bias current in dimensioned units.  The voltages displayed on the lower axis are normalized to the characteristic voltage $V_{\rm c}= I_{\rm{c} 0}R$, where $R = 0.5\,\Omega$ is the 4.2\,K value of the $c$-axis resistance per junction. The top axis gives the total voltage of the 700 IJJ stack in dimensioned units. The color scale indicates the time-averaged power of the resistive in-plane currents dissipated in the stack. One notes that there is a significant power over a wide range of bias currents, which was one of the reasons to fix the parameters of the model to the present values. The displayed IVC is obtained by sweeping the bias current from large values down to zero, i.e. it represents the outermost return branch of the (multivalued) IVC. The horizontal dashed line indicates the 50\,K critical current which would have appeared for a bias current increasing from zero. The solid horizontal lines indicate special values of the normalized bias current of 0.24 and 0.6. The lower value is located in the low-bias regime while the higher one is in the regime where a hot spot covers about half of the stack. We use these values as a starting point for our simulations, where we fix the current to the values in the ``left'' stack while varying the current through the ``right'' stack.

Figure \ref{fig:IVCcsg}(b) shows the simulated power density dissipated by in-plane currents vs. the $y$ coordinate for normalized bias currents $i_{\rm{left}} = 0.24$ and $i_{\rm{right}} = 0.235$. The stacks are coupled via the normal conducting base layer. One observes pronounced oscillations which tell that a standing wave of the electric field with 3 half waves along $y$ has been excited in both stacks. This wave profile was persistent in the whole range of bias currents we investigated in the low-bias regime. In the special case we show here, the integrated power was about $92\,\mu$W in the left stack and about $88\,\mu$W in the right stack. Figure \ref{fig:IVCcsg}(c) shows analogous data for the high-bias regime, with $i_{\rm{left}} = 0.6$ and $i_{\rm{right}} = 0.58$. The integrated power was about $52\,\mu$W in the left stack and about $41\,\mu$W in the right stack. There is a small cusp in the power densities near $y = 100\,\mu$m, left of which the temperature in the stacks has exceeded the critical temperature of 85\,K. Also in the high-bias regime the observed oscillations are persistent over the range of bias currents studied.

\begin{figure}[h]
	\includegraphics[width=\columnwidth,clip]{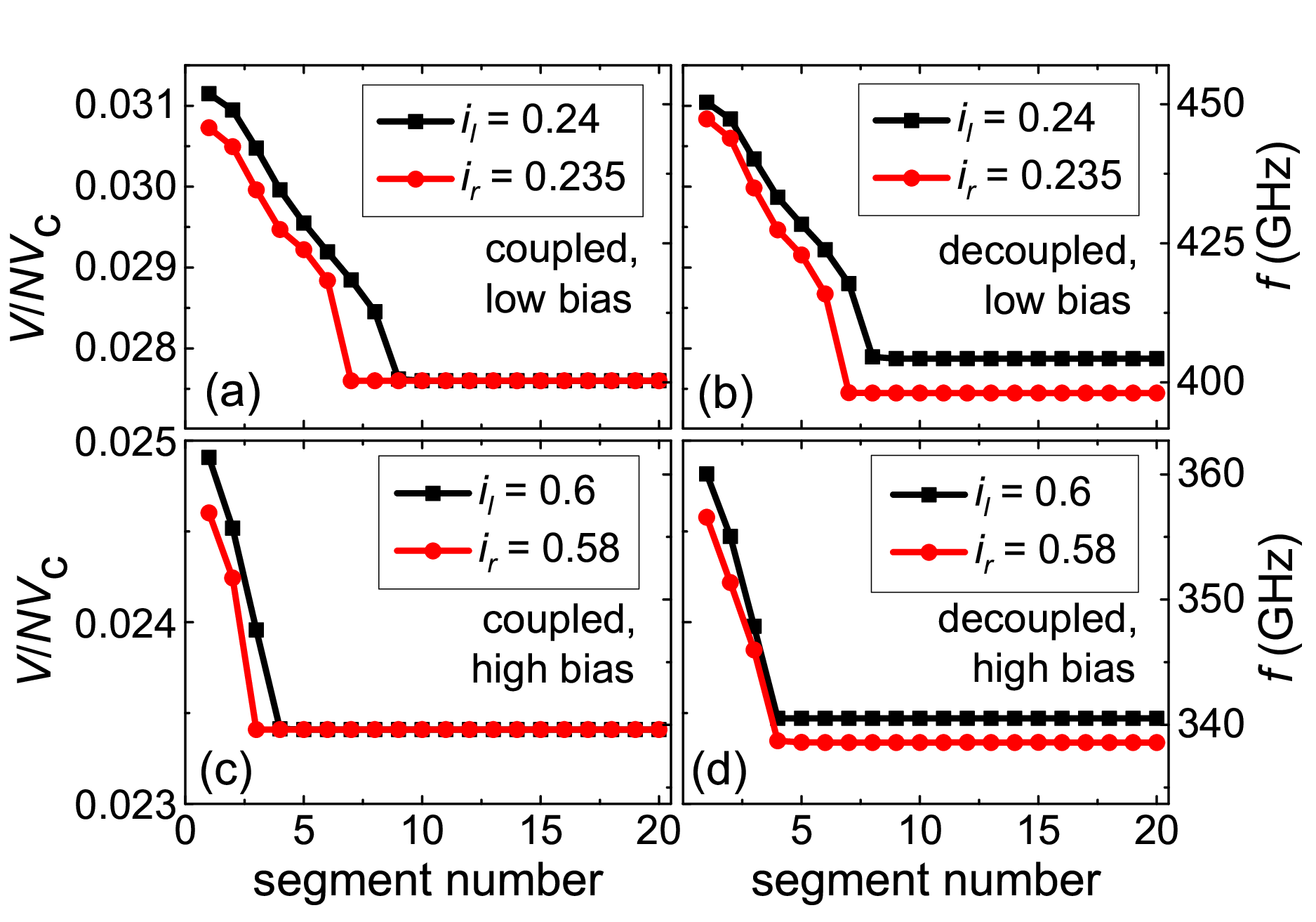}
	\caption{dc Voltage (left axes) and corresponding Josephson frequency (right axes) across individual segments vs. the segment number (bottom axes) for (a), (b) the low-bias regime and (c), (d) the high-bias regime. In (a), (c) the stacks are coupled via a normal conducting base segment, in (b), (d) the stacks are completely decoupled electrically.
	}
	\label{fig:segmentVoltage}
\end{figure}

Figure \ref{fig:segmentVoltage} displays the dc voltages across the individual segments vs. the segment number for 4 different scenarios. To obtain these voltages we average, for a given segment, the time-derivative of the Josephson phase over time and the ``cold'' parts of the segment, where the local temperature is below the critical temperature. This restriction is necessary, because in the normal-conducting part there are nonzero in-plane dc voltages which would spoil spatial averaging. In Figure \ref{fig:segmentVoltage}(a) we consider the low bias-regime, where the two stacks are coupled and biased at slightly different (normalized) currents $i$ of 0.24 and 0.235. For the left stack, biased at $i = 0.24$, the normalized dc voltage, displayed on the left axis, is the same for segments 9 to 20, while the dc voltage of segments 1 to 8 has a higher value. This shows that only a part of the segments in the stack is locked. The right axis shows the corresponding Josephson frequencies which are near 400\,GHz for the locked segments. The dc voltages across the segments of the right stack, biased at $i = 0.235$, behave similar, and on top of that, the voltages across both stacks coincide for segments 9 to 20. In our simulations we have the possibility to remove also the mutual coupling mediated by the base segment. The result is shown in Figure \ref{fig:segmentVoltage}(b). Apart from decoupling, all other conditions for the simulations remained the same, including the sequence of random numbers introduced by the noise current. The overall behavior is the same as for the coupled case, but now the segments 8 to 20 (left stack) and 7 to 20 (right stack) are locked within the individual stacks. However, the dc voltages and Josephson frequencies of the two stacks never coincide, as it should be expected for the different currents used to bias stacks\;$a$ and $b$, respectively. Figures \ref{fig:segmentVoltage}(c) and \ref{fig:segmentVoltage}(d) show analogous graphs for the high-bias regime.

\begin{figure}[h]
	\includegraphics[width=\columnwidth,clip]{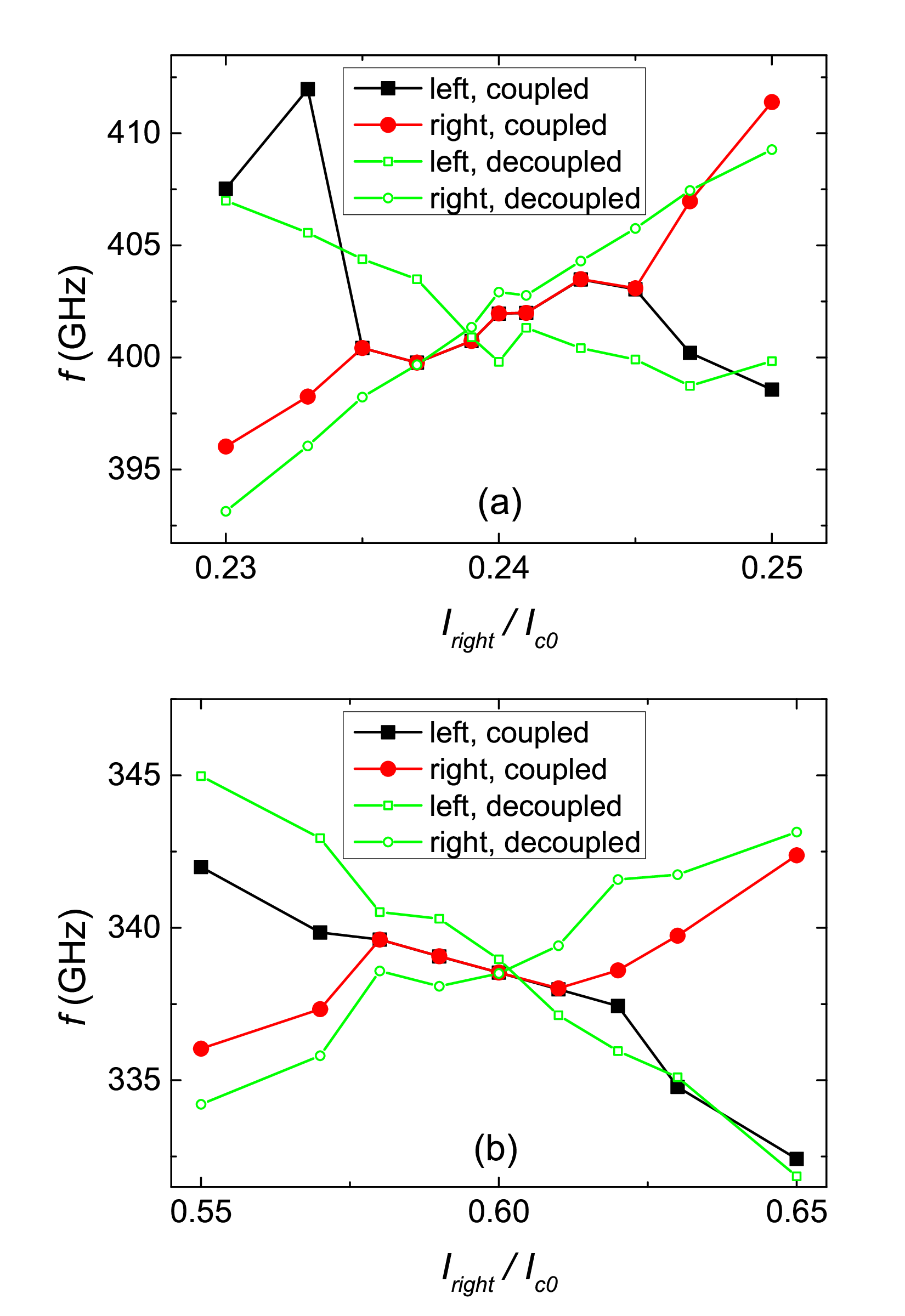}
	\caption{dc voltages across segment No. 11 in the left and right stack, converted to the Josephson frequencies, for a fixed bias current through the left stack $i_{\rm{left}} = 0.24$ in (a) and $i_{\rm{left}} = 0.6$ in (b) and a variable bias current through the right stack.
	}
	\label{fig:crossover}
\end{figure}

\begin{figure}[h]
	\includegraphics[width=\columnwidth,clip]{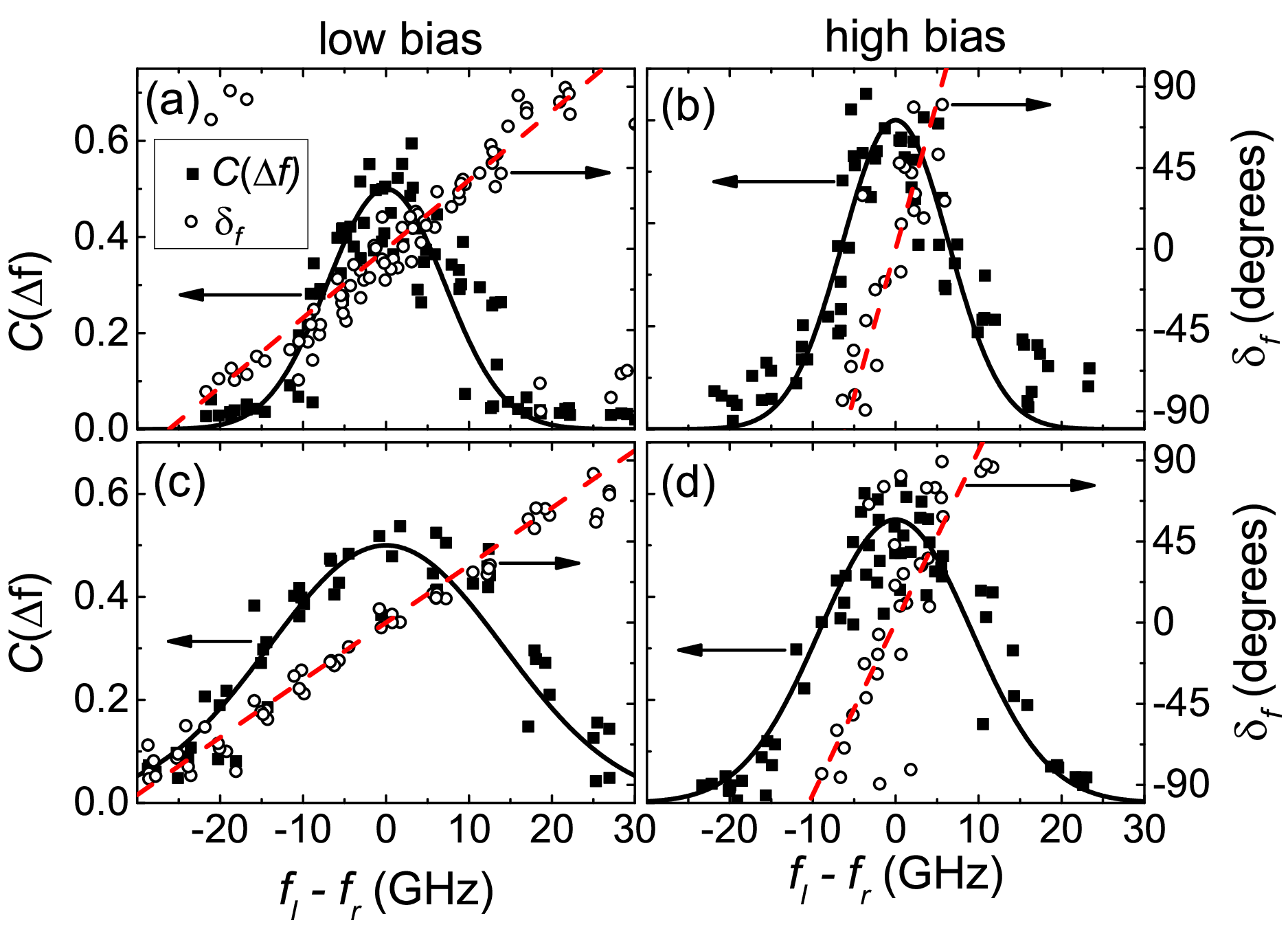}
	\caption{Correlation functions $C(\Delta f)$ (solid squares, left axes) and the phase $\delta_f$ (open circles, right axes) vs. the difference of the bare (uncoupled stacks) frequencies $f_{\rm l}-f_{\rm r}$ of, respectively the left and the right stack. (a), (b) are for a normal conducting common base segment, (c) and (d) for a fully superconducting base segment).  (a), (c): low bias regime, with the left stack biased at $i_{\rm{left}} = 0.24$; (b),(d): high-bias regime, with the left stack biased at  at $i_{\rm{left}} = 0.6$. The bath temperature is 50\,K in all cases.
	}
	\label{fig:correlationCSG}
\end{figure}

To investigate the mutual phase-lock more systematically we next fix the bias current through the left stack while sweeping the current through the right stack. Figure \ref{fig:crossover} shows the corresponding data. Somewhat arbitrarily we monitor the dc voltages across segment No. 11 for, respectively, the left and right stack. As, e.g., seen in Figure \ref{fig:segmentVoltage} this segment is inside the locked part of the individual stacks. In Figure \ref{fig:crossover} we have converted these voltages to frequency, using the Josephson relation. Figure \ref{fig:crossover}(a) is for the low-bias regime, while Figure \ref{fig:crossover}(b) is for the high-bias regime. In (a) the normalized bias current through the left stack is kept at 0.24, and the current through the right stack is varied between 0.23 and 0.25.
Solid black squares and solid red circles are for, respectively, the left and right stack and were obtained for a mutual coupling mediated by the normal conducting base segment. We see identical voltages (frequencies) of both stacks between $i_{\rm{right}} = 0.235$ and 0.245, indicating phase-lock. Outside of this regime the voltages and Josephson frequencies differ. The green open squares and open circles are for the decoupled case and yield, in the language of the Kuramoto model, the ``bare frequencies'' of the two oscillators. We see that these frequencies can differ substantially from the ``dressed frequencies'' in the coupled case. Graph \ref{fig:crossover}(b) shows the corresponding data for the high-bias regime where the left stack was biased at a normalized current of 0.6, while the current through the right stack was varied between 0.55 and 0.65. The overall behavior is similar to the low-bias case. One also notes that in both Figures \ref{fig:crossover}(a) and \ref{fig:crossover}(b) the data look noisy. In fact, it turned out that, when we repeat calculations as in Figures. \ref{fig:crossover} or \ref{fig:correlationCSG} for a \textit{different} sequence of random numbers entering via the $c$-axis noise current for otherwise the same initial conditions, one ends up with slightly different voltages. While this is unproblematic in terms of absolute voltages and/or frequencies, it is problematic for the high-resolution analysis of mutual phase-lock. The reason is that, depending on history, slightly different profiles of the local stack temperatures and the Josephson phases established and seem to be long-lived on the scale of our integration times (which however in dimensioned units are only on the order of 50\,ns and thus 7 orders of magnitude shorter than in experiment).

To address the problem, we repeated each calculation 5 times, using different sequences of random numbers. The result for the analysis of phase-lock is shown in Figure \ref{fig:correlationCSG}. Here we first defined the quantities $\cos \delta$ and $\sin \delta$, where $\delta$ is the difference of the momentary and local Josephson phases in adjacent points in the left and right stack. We then average these quantities over the cold parts of the stack and over time, to obtain the average quantities $<\cos \delta>$ and $<\sin \delta>$, which we use to define the correlation function $C(\Delta f)=\sqrt{<\cos \delta>^2+<\sin \delta>^2}$ and the phase $\delta_f=\tan^{-1}(<\sin \delta>/<\cos \delta>)$. These quantities were already introduced in the main text.  Here $\Delta f = f_{\rm b}-f_{\rm a}$ is the difference of the bare Josephson frequency, which we obtain from our simulations using decoupled stacks. 

Figures \ref{fig:correlationCSG}(a) and \ref{fig:correlationCSG}(b) show results for, respectively, the low-bias regime and the high-bias regime for the case that mutual coupling is mediated by a normal-conducting base segment. The horizontal axes display the difference in the bare frequencies of the two oscillators. For comparison, Figures. \ref{fig:correlationCSG}(c) and \ref{fig:correlationCSG}(d) show the corresponding results for the case of a superconducting base segment representing the minimal model of two mesas located on a common base crystal. In all graphs the left axes display $C(\Delta f)$ while the right axes display $\delta_f$.  Graphs \ref{fig:correlationCSG}(a) and \ref{fig:correlationCSG}(c) are for the low bias regime, where the left stack is biased at $i_{\rm{left}} = 0.24$. Graphs \ref{fig:correlationCSG}(b) and \ref{fig:correlationCSG}(d) are for the high-bias regime, with $i_{\rm{left}} = 0.6$. For each pair of bias currents through, respectively, the left and right stack we performed the simulations 5 times for different sequences of random numbers. For a given run we used the same random numbers and performed the calculations two times, first for the coupled stacks and afterwards for the decoupled stacks to obtain the bare frequencies. 

In each graph we compare $C(\Delta f)$ with a Gaussian, $C(\Delta f)=C_{\rm 0}e^{-[(f_{\rm b}-f_{\rm a})/f_{\rm C}]^2}$ (solid lines) and for the phase $\delta_f$ we use the linear relation $\delta_f=(f_{\rm b}-f_{\rm a})/f_{\rm \delta}$. For $f_{\rm C}$ we use values of 10\,GHz (a), 9\,GHz (b), 20\,GHz (c) and 13\,GHz (d), and for $f_{\rm \delta}$ we use 15\,GHz (a), 3.5\,GHz (b), 18\,GHz (c) and 6\,GHz (d). These values are at least an order of magnitude larger than the experimental ones, which may not be surprizing, because in our toy model there is no real physical gap between the two oscillators. On the other hand, Figure \ref{fig:correlationCSG} shows that the functional dependences for $C(\Delta f)$ and for $\delta_f$, plus the description as a whole, seem reasonable.  
\section{Analysis of the Lock-in detected signal }
\setcounter{figure}{0}
\label{Appendix D}
In the main text the expressions
\begin{align*}
	\label{eq:S3aAppendixD}
	\tag{\ref{eq:S3}a}
		S_{\rm X} &= \frac{P_{\rm{X0}}}{P_{\rm{Xa}}+P_{\rm{Xb}}} C(\Delta f) \cos(\delta_f + \varphi_{\rm X}) \\
		&= S_{\rm{X0}} \, C(\Delta f) \cos(\delta_f + \varphi_{\rm X})
\end{align*}

\begin{align*}
	\label{eq:S3bAppendixD}
	\tag{\ref{eq:S3}b}
	S_{\rm Y} &= \frac{P_{\rm{Y0}}}{P_{\rm{Ya}}+P_{\rm{Yb}}} C(\Delta f) \cos(\delta_f + \varphi_{\rm Y}) \\
	&= S_{\rm{Y0}} \, C(\Delta f) \cos(\delta_f + \varphi_{\rm Y})
\end{align*}
have been used to analyze the lock-in detected power enhancement of the two interfering THz signals generated by stack\;$a$ and stack\;$b$. In the following we will derive these expressions and relate them to the parameters of our experimental setup. We relate Eqs. (\ref{eq:S3}a) and  (\ref{eq:S3}b) to the local wave fronts emitted by the two stacks. To do so we first have a more precise look on the geometry and the different components of the experiment. After some general considerations we consider an incoherent signal as it has been used for the hot/cold calibration. Next, we investigate the lock-in signal detected from a coherent beam of a single stack and finally we turn to the case of interfering coherent beams.

\subsection{Geometry and general considerations}

\begin{figure}[h]
	\includegraphics[width=\columnwidth,clip]{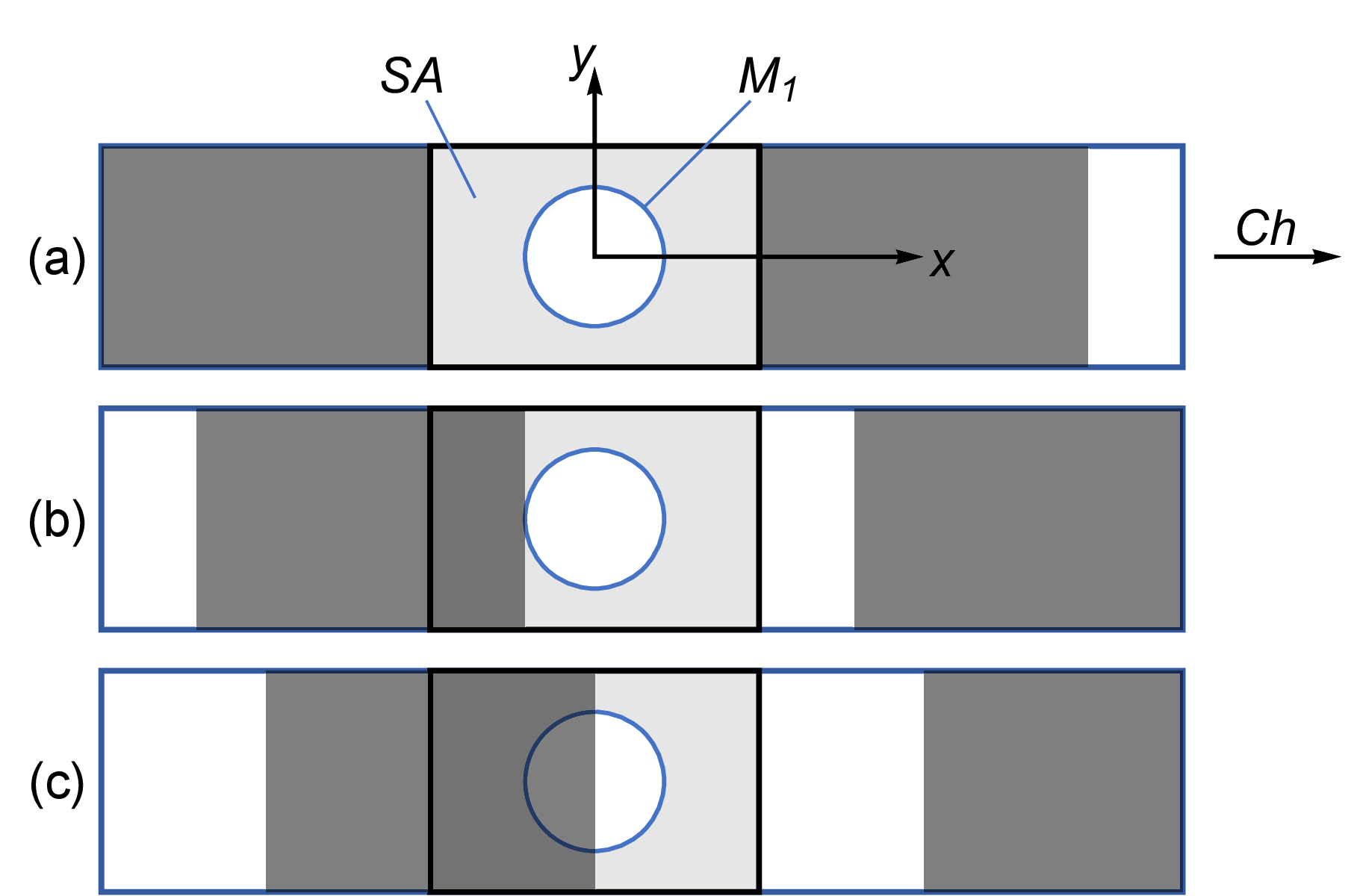}
	\caption{Projections in the plane $(x,y)$ of the chopper $(Ch)$. The coordinate system is centered at the optical axis. The chopper is sketched by alternating dark grey and white rectangles of the same size $L_c$ in $x$ direction. The center circle represents a projection of the first parabolic mirror $M_1$ onto the chopper plane. $SA$ defines the maximum signal area, defined as the open area when one of the chopper windows is centered on the optical axis. Sketches (a) to (c) are for three different times and thus chopper positions.
	}
	\label{fig:chopper}
\end{figure}

The overall geometry of the setup and the detection scheme is outlined in the main paper, c.f. Figure \ref{fig:setupTue}. 
An important component is the chopper which periodically modulates the interfering incoming light fields. Figure \ref{fig:chopper} sketches projections of various components in the plane $(x,y)$ of the chopper. The view is along the optical axis (``z-direction''). The moving chopper $(Ch)$ is sketched by a series of dark grey and white rectangles which move to the right with increasing time. This is a simplification, because the actually open and dark sections of the chopper are approximately triangles rotating around an axis which is laterally displaced from the optical axis. In other words, we approximate the action of the chopper by a periodic box-function. 

For convenience we introduce a dimensionless time $\tau=\omega_{Ch}t$, i.e. one full dark-bright period corresponds to $2\pi$. We measure spatial coordinates in units of $L_c/\pi$, where $L_c$ is the length of the subsequently open (white) and closed (dark grey) parts of the chopper in $x-$direction. $\xi=\pi x/L_c$ is the dimensionless coordinate in $x-$direction and $\zeta=\pi y/L_c$ the dimensionless coordinate in $y-$direction.
Figure \ref{fig:chopper}(a) shows the chopper at a time  when the open part of the chopper is centered on the optical axis. For this position we define a static maximum signal area $(SA)$, which is fixed to the optical axis and coincides with the open part of the chopper. In dimensionless units the $SA$ extends from $\xi=-\pi/2$ to $\xi=+\pi/2$ .  The plot also defines by the on-axis white circle the projection of the first parabolic mirror $M_1$, which shall extend from $-r$ to $r$ in $\xi-$direction, with $r \leq 2\pi$. $M_1$ defines the relevant area where the emitted radiation is collected. With respect to the full $SA$ we can simply assume that the amplitude of the emitted radiation is zero outside the area of $M_1$. 

The right edge of the open part of the chopper is ahead in time relative to its center by a time difference $\Delta \tau = \pi/2$ and thus reaches a certain spatial location $\xi$ on the $SA$ and $M_1$ at $\tau=\pi/2 - \xi$. Correspondingly, the left edge of the open part of the chopper is by a time difference $\Delta \tau =-\pi/2$ behind the center. Figure \ref{fig:chopper}(b) is for a time $\tau=\pi/2-r$, when the left edge has reached $M_1$. Figure \ref{fig:chopper}(c) is for $\tau=\pi/2$ when half of the $SA$ and $M_1$ are shadowed by the chopper. 

\begin{figure}[h]
	\includegraphics[width=0.9\columnwidth,clip]{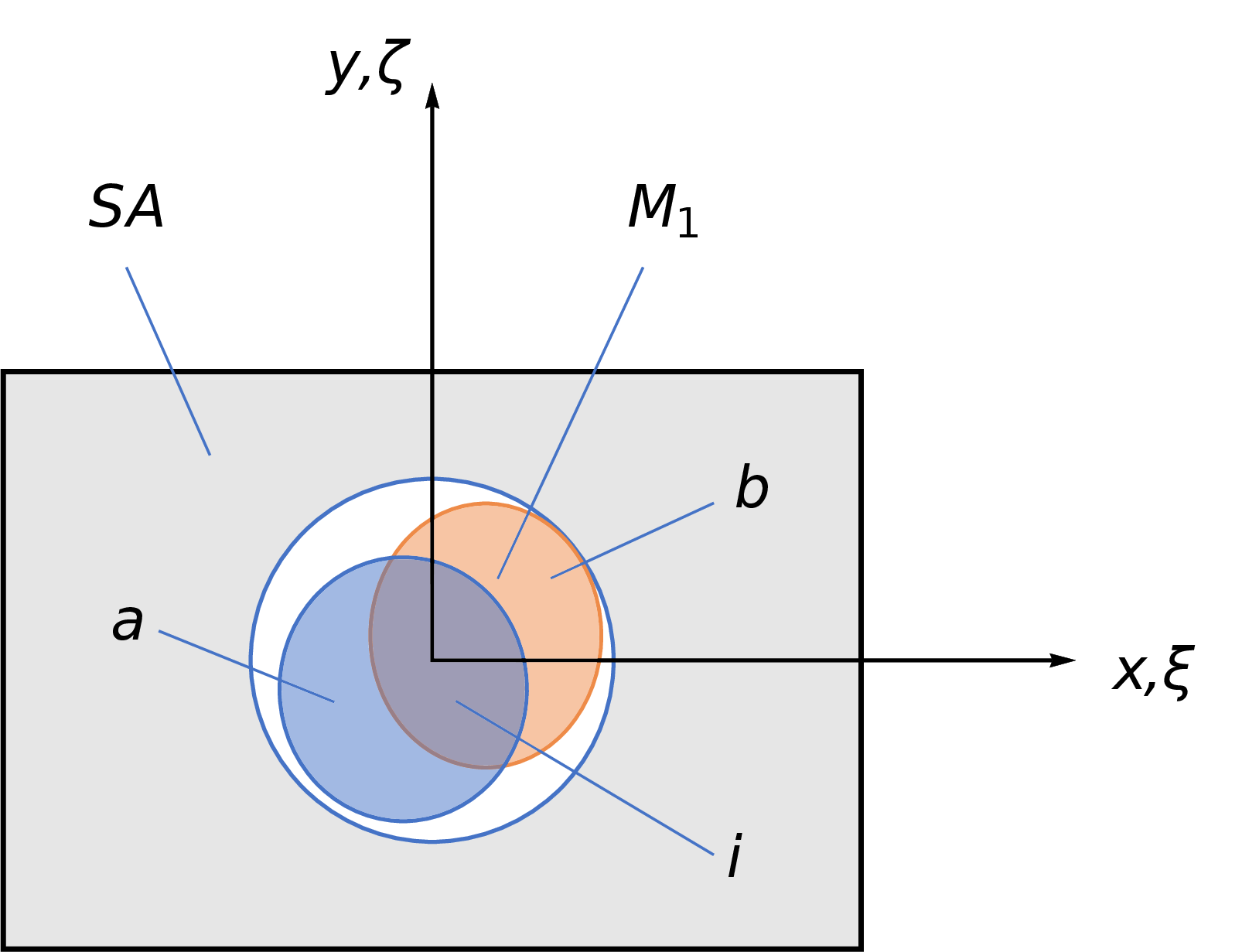}
	\caption{Maximum signal area $(SA)$, projection of the first parabolic mirror $(M_1)$ and regions illuminated by stacks\;$a$ and $b$, plus an overlap area $i$ in the plane of the chopper.
	}
	\label{fig:SA}
\end{figure}

Figure \ref{fig:SA} displays a closer look on $SA$ and $M_1$ in the presence of the radiation fields emitted by the stacks\;$a$ and $b$. These fields will be at least partially off-axis and presumably have some overlap region $i$. For the hot/cold calibration using thermal radiation we will assume that $M_1$ is illuminated concentrically. 

In all situations we can characterize the incoming signal by a complex light field propagating (approximately) along the optical axis,
\begin{equation}
	\label{eq:E1}
	E(\xi,\zeta,t)=E_0(\xi,\zeta)\cdot e^{i[\varphi(\xi,\zeta)-\omega t]}
\end{equation}
in the plane of the chopper.  Having passed the lens, the phase-field $\varphi(\xi,\zeta)$ may vary strongly in an unknown way, as discussed in the main paper. Next, the light field described by Eq. (\ref{eq:E1}) is periodically opened and shadowed by the chopper, leading to a time modulated light field 
\begin{equation}
	\label{eq:E2}
	E(\tau)=E_0(\xi,\zeta)\cdot L(\xi,\tau) \cdot e^{i[\varphi(\xi,\zeta)-\omega t]},
\end{equation}
where $L(\xi,\tau)$
parametrizes the size and position of the open part of $M_1$. For an opening chopper window $L(\xi,\tau)$ can be written as
\begin{align*}
	\label{eq:La}
	\tag{D3a}
	L(\xi,\tau)=0 & \\ & \textrm{ for } -\pi \leq \tau < \tau_1 = -\pi/2-r \\ & \textrm{ ``$M_1$ not reached yet''}
\end{align*}

\begin{align*}
	\label{eq:Lb}
	\tag{D3b}
	L(\xi,\tau)=\Theta(\xi+r)&-\Theta(\xi-\tau-\pi/2) \\ &\textrm{ for } \tau_1 < \tau < \tau_2 = -\pi/2+r \\ &\textrm{ ``open part of $M_1$ grows''}
\end{align*}

\begin{align*}
	\label{eq:Lc}
	\tag{D3c}
	L(\xi,\tau)=\Theta(\xi+r)&-\Theta(\xi-r) \\ &\textrm{ for } \tau_2 \leq \tau \leq 0 \textrm{ ``$M_1$ open''}
\end{align*}
For $0<\tau \leq \pi$ the window closes in reverse order, i.e. we have expression (D3) for $0 \leq \tau \leq \tau_3 = \pi/2-r$, 
\begin{align*}
	\label{eq:Ld}
	\tag{D3d}
	L(\xi,\tau)=\Theta(\xi-\tau+\pi/2)&-\Theta(\xi-r) \\ &\textrm{ for } \tau_3 < \tau < \tau_4=\pi/2+r
\end{align*}
and $L(\xi,\tau)=0$ for $\tau_4 < \tau \leq \pi$.

In the expressions above we have ignored the thermal radiation also appearing when the chopper window has closed. This part contributes to the background, which is partially taken care of by the lock-in detection scheme and also by our background subtraction procedures. 

Finally, behind the chopper the (still propagating but now chopped) light field $E(\tau)$ propagates to the bolometer, building up (unknown) additional phase differences and spatial correlations.
Quantitatively we may describe the light field arriving in the bolometer plane by $E_{\rm{bol}}(\xi,\zeta,\tau)$, where the electric field at a given point $(\xi,\zeta)$ results from an integration of the form
\setcounter{equation}{3}
\begin{equation}
	\label{eq:E3}
	E_{\rm{bol}}(\xi,\zeta,\tau,t)=\int \mathrm{d}\xi' \mathrm{d} \zeta' \cdot E(\tau) \cdot K(\xi-\xi',\zeta-\zeta'),
\end{equation}
with some (complex-valued) spatial correlation function $K(\xi-\xi',\zeta-\zeta')$. $E_{\rm{bol}}(\xi,\zeta,\tau,t)$ may be writen as
\begin{equation}
	\label{eq:E4}
	E_{\rm{bol}}(\xi,\zeta,\tau,t) = E_{\rm{bol}\,0}(\xi,\zeta,\tau) e^{i[\varphi_0+\varphi_1(\xi,\zeta, \tau)-\omega t]},
\end{equation}
where we have separated out a global phase $\varphi_0$.

We should then integrate the intensity $E^2_{\rm{bol},0}(\xi,\zeta,\tau)$ over the area of the bolometer to yield a final output $P(\tau)$ detected by the lock-in. Given a proper calibration of the common phase $\varphi_{\rm c}$ in the lock-in channels X and Y, the lock-in outputs, after multiplication with, respectively, cos$(\tau)$ (X channel) and sin$(\tau)$ (Y channel)  project out the symmetric (X) and antisymmetric (Y) part of the incoming signal $P(\tau)$.

\subsection{Analysis of various light fields}
We consider three cases, namely the case of thermal radiation during the hot/cold calibration measurements, the case of coherent beams of emitters\;$a$ and $b$ which are mutually unsynchronized, and finally the case of mutually synchronized emitters\;$a$ and $b$.

\subsubsection{Hot/cold calibration}
Here, the (thermal) light field is fully incoherent and contains many emission frequencies. The correlation function $K(\xi-\xi',\zeta-\zeta')$ is a delta-function in both $\xi$ and $\zeta$. From Eqs. (D3) we find that, assuming that all optical components are well aligned to the optical axis, the response $P(\tau)$  is symmetric in $\tau$, if the intensity distribution of the (thermal) light field in the chopper plane was spatially symmetric (i.e., relative to coordinates $\xi$ and $\pi$) on the mirror $M_1$.  This (ideally) leads to a finite signal in the X channel, and zero signal in the Y channel.

\subsubsection{Single oscillator emitting coherently}
$E_{\rm{bol}}(\xi,\zeta,\tau)$ in general contains interference terms and may have a complicated spatial distribution. The light field $E(\xi,\zeta,t)$ in the chopper plane is not necessarily symmetric with respect to the optical axis, c.f. Figure \ref{fig:SA}, and on top of that the spatial correlations built up by $K(\xi-\xi',\zeta-\zeta')$ may break original symmetries. Thus, we expect that the lock-in detected signal  $P(\tau)$ does not appear exclusively in the X channel, which is the case in our experiments. This can be seen in the in Figure \ref{fig:fitResults2x2} of Appendix \ref{Appendix B} showing the four reconstructed curves of the emission powers $P_{\rm{a,X}}$, $P_{\rm{b,X}}$, $P_{\rm{a,Y}}$ and $P_{\rm{b,Y}}$  vs. the respective emission frequency,  where roughly 20\% of the signal appears in the Y channel. One also notes that for stack\;$a$ the shape of the curves is quite different in the X and the Y channel. For stack\;$b$ the differences are less pronounced but still obvious. This rules out that the Y channel is just a low-signal replica of the X channel, which would have happened if the Y signal would have a nonzero signal just due to an incorrect setting of the lock-in reference phase. 

\subsubsection{Mutually synchronized oscillators $a$ and $b$}
In the plane of the bolometer we have a light field of the form
\begin{equation}
	\label{eq:E5}
	\begin{split}
	E_{\rm{bol}}(\xi,\zeta,\tau,t) 
	&= E_{\mathrm{bol}\,0,a}(\xi,\zeta,\tau) e^{i[\varphi_{0a}+\varphi_{1a}(\xi,\zeta,\tau)-\omega_a t]} \\
	&+ E_{\mathrm{bol}\,0,b}(\xi,\zeta,\tau) e^{i[\varphi_{0b}+\varphi_{1b}(\xi,\zeta,\tau)-\omega_b t]} ,
	\end{split}
\end{equation}
The intensity $E_{\rm{bol}}E^*_{\rm{bol}}$ is given by
\begin{equation}
	\label{eq:E6}
	\begin{split}
		&E_{\rm{bol}}(\xi,\zeta,\tau)E^*_{\rm{bol}}(\xi,\zeta,\tau)
		= E^2_{\rm{bol}\,0,a} + E^2_{\rm{bol}\,0,b} \\
		&+ 2E_{\mathrm{bol}\,0,a} + E_{\mathrm{bol}\,0,b} \cos[\delta +\Delta \varphi(\xi,\zeta,\tau)],
	\end{split}
\end{equation}
with the oscillator phase difference $\delta$, c.f. Appendix \ref{Appendix C}, and the geometric phase difference $\Delta \varphi(\xi,\zeta,\tau)= \varphi_b(\xi,\zeta,\tau) - \varphi_a(\xi,\zeta,\tau)$. Here we omitted the arguments in the amplitude functions for simplicity. The last term on the right-hand side is the interference term. 
After time-averaging over the fluctuations of $\delta(t)$ one obtains
\begin{equation}
	\label{eq:E7}
	\begin{split}
		&E_{\rm{bol}}(\xi,\zeta,\tau,t)E^2_{\rm{bol}}(\xi,\zeta,\tau,t)
		= E^2_{\rm{bol}\,0,a} + E^2_{\rm{bol}\,0,b} \\
		&+ 2E_{\mathrm{bol}\,0,a}  E_{\mathrm{bol}\,0,b} <\cos[\delta +\Delta \varphi(\xi,\zeta,\tau)]>,
	\end{split}
\end{equation}
with expressions $C(\Delta f)=\sqrt{<\cos \delta>^2+<\sin \delta>^2}$, $\delta_f=\tan^{-1}(<\sin \delta>/<\cos \delta>)$ and $\Delta f = f_{\rm b}-f_{\rm a}$. given in the main paper and discussed in Appendix \ref{Appendix C} the time-averaged cosine gives
\begin{equation}
	\label{eq:cosine1}
	 <\cos[\delta +\Delta \varphi(\xi,\zeta,\tau)]> = C(\Delta f) \cos[\delta_f +\Delta \varphi(\xi,\zeta,\tau)].
\end{equation}
Next, Eq. (\ref{eq:E7}) should be integrated over the bolometer area yielding $P(\tau)$, to be multiplied with, respectively, cos$(\tau)$ and sin$(\tau)$ and integrated over time.  Expanding the cos Term in (\ref{eq:E7}) we obtain
\begin{equation}
	\label{eq:cosine2}
	\begin{split}
		&2\sqrt{2} E_{\mathrm{bol}\,0,a}  E_{\mathrm{bol}\,0,b} C(\Delta f) \\
		&\cdot \{\cos(\delta_f) \cos[\Delta\varphi(\xi,\zeta,\tau)]-\sin(\delta_f) \sin[\Delta\varphi(\xi,\zeta,\tau)] \}
	\end{split}
\end{equation}
and from there after integration over the bolometer area and time one finds for the lock-in detected interference term in the X channel
\begin{equation}
	\label{eq:P1}
	P_{\mathrm{X}ab} = P_{\rm X, cos}C(\Delta f)\cos(\delta_f) - P_{\rm X, sin}C(\Delta f)\sin(\delta_f),
\end{equation}
which can be reformulated as
\begin{equation}
	\label{eq:P2}
	\begin{split}
		&P_{\mathrm{X}ab} = P_{\rm X, 0}C(\Delta f)\cos(\delta_f + \varphi_{\rm X}) \textrm{, with }\\
		&P_{\rm X, 0}=\sqrt{P^2_{\rm X, cos} + P^2_{\rm X, sin}} \textrm{ and } \varphi_{\rm X}=\tan^{-1}\left(\frac{P_{\rm X, sin}}{P_{\rm X, cos}}\right).
	\end{split}
\end{equation}
Similarly, we obtain for the Y channel
\begin{equation}
	\label{eq:P3}
	\begin{split}
		&P_{\mathrm{Y}ab} = P_{\rm Y, 0}C(\Delta f)\cos(\delta_f + \varphi_{\rm Y}) \textrm{, with }\\
		&P_{\rm Y, 0}=\sqrt{P^2_{\rm Y, cos} + P^2_{\rm Y, sin}} \textrm{ and } \varphi_{\rm Y}=\tan^{-1}\left(\frac{P_{\rm Y, sin}}{P_{\rm Y, cos}}\right).
	\end{split}
\end{equation}
The overall response in the two channels can be written as
\begin{equation}
	\label{eq:P4}
	\tag{X channel, D14a}
	\begin{split}
	P_{\mathrm{X}} = P_{\rm X, a} + P_{\rm X, b} + P_{\rm X, 0} &C(\Delta f)\cos(\delta_f + \varphi_{\rm X}) 
	\end{split}
\end{equation}
\begin{equation}
	\label{eq:P5}
	\tag{Y channel, D14b}
	\begin{split}
		P_{\mathrm{Y}} = P_{\rm Y, a} + P_{\rm Y, b} + P_{\rm Y, 0} &C(\Delta f)\cos(\delta_f + \varphi_{\rm Y}) 
	\end{split}
\end{equation}
After determining the background relative to the interference term, as outlined in Appendix \ref{Appendix B}, and division by the background one obtains 
\begin{equation}
	\label{eq:S4}
	\tag{D15a}
	\begin{split}
		S_{\mathrm{X}} &= \frac{P_{\mathrm{X}, 0}}{P_{\mathrm{X}, a} + P_{\mathrm{X}, b}} C(\Delta f)\cos(\delta_f + \varphi_{\rm X}) \\ 
		&= S_{\mathrm{X}, 0} C(\Delta f)\cos(\delta_f + \varphi_{\rm X})
		\textrm{\qquad (X channel)}
	\end{split}
\end{equation}
\begin{equation}
	\label{eq:S5}
	\tag{D15b}
	\begin{split}
		S_{\mathrm{Y}} &= \frac{P_{\mathrm{Y}, 0}}{P_{\mathrm{Y}, a} + P_{\mathrm{Y}, b}} C(\Delta f)\cos(\delta_f + \varphi_{\rm Y}) \\ 
		&= S_{\mathrm{Y}, 0} C(\Delta f)\cos(\delta_f + \varphi_{\rm Y})
		\textrm{\qquad (Y channel)}
	\end{split}
\end{equation}
which are the functions used in the main text.

Quantities $P_{\mathrm{X}, a}$, $P_{\mathrm{X}, b}$, $P_{\mathrm{Y}, a}$ and $P_{\mathrm{Y}, b}$ can be obtained from the experimental data via analysis of the background as outlined in Appendix \ref{Appendix B}. Functions $P_{\mathrm{X}, 0}$ and $P_{\mathrm{Y}, 0}$ are not accessible independent of $C(\Delta f=0)=C_0$. However, we note that $P_{\mathrm{X}, 0}$ and $P_{\mathrm{Y}, 0}$ result from spatial integrations over the terms $E_{\mathrm{bol}\,0,a} E_{\mathrm{bol}\,0,b} \cos[\Delta\varphi(\xi,\zeta)]$ and $E_{\mathrm{bol}\,0,a} E_{\mathrm{bol}\,0,b} \sin[\Delta\varphi(\xi,\zeta)]$ in Eq. (\ref{eq:cosine2}). They can be very small if the cos and sin functions change sign often over the bolometer area. As a consequence, small values of the products $S_{\mathrm{X}, 0} C_0$ and $S_{\mathrm{Y}, 0} C_0$ do not imply, that the maximum value of $C_0$ is small. Thus, we can get information on the shape and the width of $C(\Delta f)$ but not on its magnitude $C_0$. More details are given in the main text.

\end{document}